\documentclass[universe,article,accept,moreauthors]{Definitions/mdpi}

\firstpage{1} 
\pubvolume{1}
\issuenum{1}
\articlenumber{0}
\pubyear{2026}
\copyrightyear{2026}
\externaleditor{Academic Editor: Firstname Lastname}
\datereceived{14 January 2026} 
\daterevised{17 March 2026} 
\dateaccepted{23 March 2026} 
\datepublished{ } 
\colorlet{red}{black}

\usepackage{mathtools}
\usepackage{gensymb}
\usepackage{tikz}

\usepackage{graphicx}	
\usepackage{amsmath}	
\usepackage{amssymb}	

\newcommand{\myarrow}{\tikz\draw[thin,black,-latex] (1,4.5ex) -- ++(0,-2.05ex) -- +(4.5ex,0); \;}

\usepackage[normalem]{ulem}

\Title{The `Forgotten' Neutrons: Implications for the Propagation \linebreak of High-Energy 
Cosmic Rays in Magnetized Astrophysical \linebreak and Cosmological Structures}

\Author{Ellis R. Owen $^{1,2,}$*\orcidA{}, 
Kinwah Wu $^{3,4,5,}$*\orcidB{},  
Yoshiyuki Inoue $^{2,4,6}$\orcidC{}, 
Tatsuki Fujiwara $^{2}$\orcidD{}, 
Qin Han $^{3,2}$\orcidE{} \linebreak and 
Hayden P. H. Ng $^{7}$\orcidF{} 
 }

\AuthorNames{Ellis R. Owen, Kinwah Wu, Yoshiyuki Inoue, Tatsuki Fujiwara, 
Qin Han, Hayden P. H. Ng}

\address{%
$^{1}$ \quad Astrophysical Big Bang Laboratory (ABBL), RIKEN Pioneering Research Institute (PRI), 2-1 Hirosawa, \linebreak  Wak\={o} 351-0198, Saitama, Japan\\ 
$^{2}$ \quad Theoretical Astrophysics, Department of Earth and Space Science, Graduate School of Science, The University of Osaka, 1-1 Machikaneyama, Toyonaka 560-0043, Osaka, Japan \\  
$^{3}$ \quad  Mullard Space Science Laboratory, University College London, Holmbury St.~Mary, Surrey RH5 6NT, UK\\
$^{4}$ \quad Kavli Institute for the Physics and Mathematics of the Universe (WPI), 
University of Tokyo Institutes for Advanced Study, The University of Tokyo, Kashiwa 277-8583, Chiba, Japan \\ 
$^{5}$ \quad Research School of Astronomy and Astrophysics, 
  Australian National University, 
  \linebreak Canberra, ACT 2611, Australia \\ 
$^{6}$ \quad RIKEN Center for Interdisciplinary Theoretical and Mathematical Sciences (iTHEMS), 2-1 Hirosawa, \linebreak Wak\={o} 351-0198, Saitama, Japan \\
$^{7}$ \quad Department of Physics and Astronomy, University College London, Gower Street, London WC1E 6BT, UK
}

\corres{Correspondence:  
ellis.owen@riken.jp (E.R.O.); 
kinwah.wu@ucl.ac.uk (K.W.) }

\abstract{Cosmological filaments, galaxy clusters, and galaxies are magnetized reservoirs of cosmic rays (CRs). 
The exchange of CRs across these structures is usually modeled assuming that they remain charged and magnetically confined. 
At high energies, hadronic interactions can convert CR protons to neutrons. This physics is routinely included in air-shower and ultra-high-energy (UHE) CR propagation Monte Carlo simulations used for composition studies but is rarely treated explicitly in propagation models of CR transport and exchange between magnetized reservoirs.  
CR neutrons are not affected by magnetic fields and can propagate ballistically over kpc-Mpc distances before decaying back into protons, with relativistic time dilation extending their effective decay length.
We show how such charged--neutral switching modifies CR confinement and escape in four representative environments: a Milky Way-like galaxy, a starburst galaxy, a galaxy cluster, and a cosmological filament. 
By solving the transport of a confined CR proton population in each structure using a diffusion/streaming propagation approach with hadronic pp and p$\gamma$ interactions, and treating neutron production and decay as a stochastic Poisson ``jump'' process, we find that neutron-mediated steps can allow additional 
CR escape from large-scale cosmological structures at energies where 
 charged-particle transport alone would predict strong CR confinement and attenuation in ambient radiation fields.  
 These effects imply a qualitative shift in how ultra-high-energy CRs are transferred from embedded sources into filaments and voids once intermediate neutron propagation is considered, with consequences for the partitioning of CRs across the large-scale structure of the Universe.}

\keyword{
Astroparticle physics; 
cosmic rays; 
magnetic fields; 
cosmological filaments; 
cosmological voids; 
(galaxy) clusters; 
galaxies;  
hadronic interactions; 
radiative processes 
} 

\makeatletter
\providecommand{\@societyowner}{}
\makeatother

\begin{document}




%
\section{Introduction}
\label{sec:introduction}

Galaxy clusters, super-clusters, 
 cosmic filaments and voids  
   are the largest structures in the Universe.  
They sit at the top of the cosmic hierarchy, 
   hosting lower-order systems 
   such as galaxies and galaxy groups. 
These large structures are permeated by magnetic fields  
   of different scales and topologies 
  \cite{Govoni2004IJMPD, Vernstrom2021MNRAS, Ryu2012SSRv, Brandenburg2023ARA&A}. 
Magnetism in galaxies has been observed  
  out to very high redshifts. 
Some galaxies exhibit ordered magnetic fields,   
  with strengths exceeding $10 \;\! \upmu$G~\cite{Mao2017NatAs, Geach2023Natur, Brandenburg2023ARA&A, Chen2024arXiv}. 
The strengths of magnetic fields in 
  clusters and super-clusters 
  are slightly lower, estimated 
  to be around $\upmu$G levels~\cite{Vacca2018Galax, Ryu2008Sci}. 
The evolutionary history of magnetism 
  in galaxy clusters and super-clusters is not yet settled.  
A common view is that their development is 
   associated with turbulence. 
This drives a small-scale dynamo 
 to amplify seed magnetic fields over timescales 
  of the order of hundreds of Myr 
  \cite{Vazza2014MNRAS, Ryu2008Sci}, 
  where the seed magnetic fields 
  may be primordial or injected by galactic feedback (for a review, see~\cite{Donnert2018SSRv}).    
The presence of cosmic rays (CRs) in galaxies 
  has been recognized for decades (for reviews, see
  \cite{Ruszkowski2023A&ARv, Owen2023Galax}).  
Like galaxies, 
  clusters and super-clusters are CR reservoirs. 
CR electrons are relativistic. 
They emit synchrotron radiation 
  when gyrating around a magnetic field. 
The presence of energetic (CR) electrons 
  in clusters is evident 
  from their diffuse radio emission 
  \cite{Brunetti2014IJMPD}. 
It has been suggested 
  that a large population of energetic hadronic CRs 
  also reside in galaxy clusters and super-clusters 
  \cite{Croston2008MNRAS, Croston2018MNRAS}  
(for recent reviews, see \cite{Ruszkowski2023A&ARv, Bourne2023Galax}).

Cosmological filaments 
   contain most of the matter 
  in the Universe 
   \cite{Calvo2010MNRAS} and are believed to have formed 
  well above redshift \emph{z} $\sim$ 4~\cite{Zhu2021ApJ}. This 
 implies that 
 they have always been the dominant mass condensates of the Universe. 
Our current understanding of 
  their dynamics and thermodynamics 
  is still primitive, 
  despite their importance and their 
  prominent position at the apex 
  of the structural hierarchy  
  of the Universe. 

Filaments are expected to 
  have weaker magnetic fields 
  than clusters and super-clusters.  
Observations indicate that 
  the strengths of their magnetic fields are typically 
  in the range $\sim$20$-$60 $\rm{ nG}$  
  (e.g., \cite{Brown2017MNRAS, Vacca2018Galax, Vernstrom2021MNRAS}) 
  in the current epoch.  
So far 
  there is no clear evidence 
  showing the evolution of cluster magnetic fields 
  over cosmological timescales 
  (e.g., \cite{Carretti2023MNRAS}).  
The origin of filament magnetic fields 
 remains an open question. 
Possible explanations 
  are intrinsic cosmological processes 
  \cite{Kandus2011PhR, Subramanian2016RPPh}, 
  astrophysical feedback 
  \cite{Bertone2006MNRAS, Vazza2017CQGra, Cen2024PNAS, Qiu2025PNAS}, 
  shocks associated with structure formation~\cite{Ryu2008Sci}, and 
  a combination of these. 

Stacking observations have indicated that 
  a substantial population of CR electrons 
 reside within filaments 
  \cite{Vernstrom2021MNRAS, Vernstrom2023SciA}. 
From a theoretical perspective, 
  filaments are also expected 
  to contain a hidden relic population of 
  hadronic CRs 
 (see \cite{Wu2024Univ}).  
These hadronic CRs 
  are injected by shocks associated with structure formation \cite{Vernstrom2023SciA, Simeon2022conf}. 
They could also be transported into the filaments   
  from the lower-ordered structures, 
  e.g., via scattering in galaxy clusters,  
  as proposed by 
  some researchers 
  (e.g., \cite{Kim2019SciA}),  
  or by advection flows in feedback 
  from galaxies 
  \cite{Owen2019MNRAS, Ptuskin2013AdSpR}. 
The long cooling times and 
  the limited interaction channels available to these hadronic CRs 
  imply that they form  
  a fossil record 
  of the non-thermal energy generation history 
  of the Universe. 

The confinement of charged CRs in filaments 
  is determined by both the strength and topology  
  of the filaments' magnetic fields \cite{Wu2024Univ}. 
The magnetic fields also scatter, and hence deflect, 
the CR particles that propagate through the inter-galactic medium~\cite{Das2008ApJ}. 
In a strong scattering regime, 
  the transport of CRs in a magnetic medium 
  is a diffusive process. 
Tangled magnetic fields in filaments 
  could have coherence on multiple length scales, 
  so they could act as energy-dependent sieves. 
In general, 
  low-energy charged CRs are trapped,  
  but high-energy charged CRs 
  could break the confinement 
  and escape from the filaments. 
The magnetic confinement 
  of charged CRs 
  can be weakened or even broken 
  through various mechanisms. 
One example is in situ acceleration, 
  which can operate in filaments 
  \cite{Vernstrom2023SciA, Simeon2022conf}, 
  clusters/super-clusters 
  \cite{Wittor2023Univ, Ruszkowski2023A&ARv} 
  and galaxies/groups 
  \cite{Paul2017MNRAS, Owen2023Galax}, 
  through which CR particles 
  can acquire sufficient energies 
  to exceed the confinement threshold. 
Other processes, such as  
  cross-field diffusion~\cite{Xu2013ApJ}, 
  mechanical advection in bulk astrophysical flows 
  \cite{Owen2019MNRAS, Ptuskin2013AdSpR} 
  or tapered or folded magnetic field topologies 
  \cite{Wu2024Univ}, 
  can also facilitate CR escape.  

These scenarios implicitly assume that 
   the CR particles are all charged. 
However, CR particles can also be charge neutral. 
In certain contexts, 
  these neutral CRs can make up a substantial fraction of the overall CR population,  
  and they have significant influence 
  on the collective behavior of energetic CRs 
  as a population. 
For instance, CR neutrons, 
  produced by hadronic interactions 
  initiated by CR protons,  
  are 
  not affected by 
  magnetic fields. 
Relativistic time dilation 
  can extend CR 
  neutrons' rest-frame 
  decay time (of $\sim$10 min) 
  by orders of magnitude. 
Energetic neutrons can traverse long distances 
  before subsequently 
  decaying into protons, electrons and electron anti-neutrinos. 
A 200 PeV neutron 
  can travel across a large fraction of a galaxy 
  practically undisturbed;   
  a 200 EeV neutron  
  can easily cross a galaxy cluster without decaying. 
The temporary conversion of 
  CR protons to CR neutrons 
  thus allows very energetic CR baryons as a population    
  to break magnetic confinement  
  in galaxies, filaments and even clusters/super-clusters. 
  
The role of neutron escape in cosmic-ray transport 
 has been addressed in some source-specific studies, 
 and an example of such is the investigation 
 of the effects of neutron escape 
 on particle spectra of microquasar jets 
 \cite{Escobar2021A&A}.   
In this study, 
  we further extend 
  the previous studies, 
  in that we assess 
  the impacts of neutron--proton conversion on 
  confinement and confinement breaking of cosmic rays 
  within and between magnetized media and reservoirs,   
  and demonstrate the necessity of explicit consideration 
  of the switching between the diffusion and streaming modes 
  in the treatment of cosmic-ray transport. 
More specifically, 
  we investigate cosmic-ray transport 
  in the presence of proton--neutron conversion 
  and quantify
  how the conversion process 
  facilitates ultra-high-energy charged protons 
  to break magnetic confinement 
  in the hierarchy 
  of galaxies, galaxy clusters, and \mbox{cosmological filaments. } 
  
Throughout this work, 
the treatment of particle transport and escape considers a temporary charge-neutral phase. It is not 
neutron-specific physics. The results obtained here are applicable 
to anti-neutrons and other neutral hadrons.  
The focus of neutrons in this work 
  is a representative example  
  of the neutral-nucleon channel to demonstrate the relevant effects. 
In Section~\ref{sec:sec2}, 
 we describe the processes 
 governing the propagation and survival 
 energetic hadronic particles  
 in these structures, 
 and the methodology that we employ 
 to model neutron-assisted confinement breaking. 
In Section~\ref{sec:sec3}, 
  we demonstrate 
  how confinement breaking 
  via neutron production 
  alters the distribution of CRs 
  in galaxies, clusters/super-clusters  
  and filaments. 
In Section~\ref{sec:sec4}, 
  we discuss 
  the implications of our findings 
  in terms of CR transfer 
  in the ecosystems of  
  galaxies, clusters/super-clusters 
  and filaments. 
Concluding remarks and a summary 
  of this study 
  are presented 
  in Section~\ref{sec:sec5}. 

\section{CR Processes in Astrophysical and Cosmological Environments}
\label{sec:sec2}

\subsection{Hadronic Interactions of CR Baryons} 
\label{sec:hadronic_interctions}

Energetic CR baryons interact with ambient matter, 
   which is predominantly comprised of hydrogen, 
   via pp/np processes.  
{In the pp process, one class of  
  interactions for neutron production 
  (i.e., pp $\longrightarrow$ n+$\{X\}$, 
  where $\{X\}$ represents  
  a collection of various hadronic and leptonic particles) 
  is the diffractive dissociation process,  
  where resonant particles 
  can be produced when energies at 
   the center-of-momentum frame are sufficiently high  
(for discussions on the role of resonances in pp interactions, 
   see \citep{Tsushima1994PhLB,Xie2007PhLB,Wang2015PhRvC};  
   for discussions on particle multiplicities 
   and the energetics of the products of pp interactions, see 
  \citep{Grosse-Oetringhaus2010JPhG,Beggio2020EPJC}).  
Energetic neutrons can be produced directly 
  as secondary baryons or through resonance decays. 
Neutrons can initiate pion production through 
  np processes, analogous to the pp interaction, 
  if they do not decay into protons 
  during their propagation via $\beta^-$ decay:  
  \begin{align}
  {\rm n} \longrightarrow {\rm p}{\rm e}^{-}\bar{\nu}_{\rm e} . 
  \end{align} 
  
Another    
 class of interactions that 
 produce neutrons from energetic protons 
  is the one-pion exchange process: 
\begin{align} 
 {\rm p} +{\rm p} \longrightarrow {\rm n}{\rm p}\pi^+ . 
\end{align} 

In this process, the neutron can retain most of the energy 
  of the proton 
 (see \cite{Khoze2017PhRvD} for the comparison  
   of the two neutron-production processes).}  
  

CR baryons  
 can also interact with 
 radiation fields via p$\gamma$/n$\gamma$ interactions, 
 but generally the threshold energies are much higher,  
 as the radiation field associated 
 with large-scale astrophysical structures  
 are below keV energies. 
In p$\gamma$ processes\endnote{In addition 
   to pion production, 
   the p$\gamma$ process would lead to 
  Bethe--Heitler leptonic pair-production:   
 \begin{align}
     {\rm p}' + \gamma 
     \rightarrow {\rm p} + l^+ + l^- \ , 
 \end{align} 
 which practically operates as a continuous cooling process for the CR proton.}, 
 the dominant channels are resonant single-pion production, 
 direct single-pion production, and multiple-pion production (see \citep{Mucke1999PASA}).  
Resonant single-pion production occurs through the formation of $\Delta^{+}$ particles, which decay through two major channels. 
Both of these channels produce charged and neutral pions, secondary protons and neutrons: 
\begin{align}%
\label{eq:pg_int}%
 {\rm p}+  \gamma \longrightarrow 
	\begin{cases}%
	  \;  {\rm p} \pi^0 \longrightarrow {\rm p}\;  2\gamma				\\[0.5ex]%
	  \;  {\rm n} \pi^+ \longrightarrow {\rm n}\;\! \mu^+ \nu_{\upmu}		\\%
		\hspace{5.5em} \myarrow {\rm e}^+ \nu_{\rm e} \bar{\nu}_{\upmu}%
	\end{cases} .   
\end{align}%

The corresponding channel for neutrons is 
  the n$\gamma$ process: 
\begin{align}%
\label{eq:ng_int}%
 {\rm n} + \gamma \longrightarrow 
		\begin{cases} %
	\; 	{\rm p}\;\! \pi^-  \longrightarrow 
    {\rm p}\;\! \upmu^- \bar{\nu}_{\upmu}		\\ %
		\hspace{5.5em} \myarrow {\rm e}^- {\bar\nu}_{\rm e} {\nu}_{\upmu}%
    \\  
		\; 	{\rm n}\;\! \pi^0 \longrightarrow {\rm n}\;  2\gamma	  
		\end{cases} .   %
\end{align}%

\subsection{Charged Particle Propagation in Magnetized Media}
\label{sec:charged_prop}

Charged CR particles 
  are deflected 
  in the presence of a magnetic field.  
The extent of these deflections for 
 a CR particle,  
 of charge $Ze$ and mass $m$, 
 in a magnetic field of strength 
 $|{\boldsymbol B}|$  
 is characterized by its Larmor radius   
\begin{align}  
r_{\rm L} & 
   = 3.12\times 10^{15}\ \gamma \beta 
   \left(\frac{\sin \theta}{Z}\right)     
   \left(\frac{m}{m_{\rm p}}\right) 
   \left(\frac{|{\boldsymbol B}|}{\rm nG} \right)^{-1} \ {\rm cm}
   ,   
\end{align}  
where ${\rm m}_{\rm p}$ is the proton mass, 
  $\gamma  = (1- \beta^2)^{-1/2}$ 
  is the Lorentz factor,  
  $\beta$ is the velocity of the particle 
  (normalized to the speed of light $c$) 
  and $\theta = \cos^{-1} 
({{\boldsymbol \beta} \cdot {\boldsymbol B}}/{|\boldsymbol B|})$ is the pitch angle 
  (the angle between the direction of the particle's propagation and the local magnetic field vector).  
For a system 
  with a uniform magnetic field, 
  the Larmor radius 
  sets the critical condition for 
  the confinement of charged CRs. 
A CR particle is trapped 
  if its Larmor radius 
  is smaller than the size of the system.  
In reality, magnetic fields 
  in astrophysical systems are not uniform. They are  
 tangled, so charged CR particles are scattered 
  instead of being trapped. 
In the strong scattering regime, 
  CR particles propagate across 
  an astrophysical system by diffusion. 

For a CR nucleon 
 with a mass-to-charge ratio $A/Z$ propagating through a magnetized medium, 
  its confinement 
  can be assessed 
  using a parameter $\zeta_{\rm L}$, 
  defined as 
\begin{align}   
\zeta_{\rm L} \equiv 
  \frac{1}{2\pi{\mathcal D}} 
  \int_{2\pi} {\rm d} \Omega \ \ r_{\rm L} 
\end{align} 
 (see \cite{Wu2024Univ}), which is a measure of whether or not a charged nucleon would be confined in a domain of extent
 ${\mathcal D}$ where the magnetic field  
  has a coherent structure 
  and a characteristic field strength 
  $|{\boldsymbol B}|$.  
The particle is confined 
  when $\zeta_{\rm L} < 1$, 
  and it will break the confinement 
  when $\zeta_{\rm L} > 1$.  
When magnetic confinement holds, 
  the (macroscopic) propagation 
  of the CR is strongly coupled 
  to the medium.  
The CR propagation is 
  then well described locally 
  as a diffusion process, 
  with advection operating in bulk 
  flows of the entraining fluid. 

A full description 
 of charged CR propagation  
  is required to properly account 
  for both ballistic streaming 
  and non-ballistic transport. 
In the high-energy 
 or weakly magnetized regimes, 
 CR particles stream ballistically 
 at a speed $|\boldsymbol{v}_s| \sim c$. 
In confining structures, where $\zeta_{\rm L} < 1$, particles transition to non-ballistic transport. 
In diffusive propagation, charged particles are well-described by pitch-angle-averaged diffusion, characterized by a diffusion coefficient tensor $\boldsymbol{K}$. Within confining structures, bulk flows (of velocity $\boldsymbol{u}$) of the medium 
can advect CR particles as they diffuse. 
Additionally, the propagation of CRs along ordered magnetic topologies can introduce drift motions with velocity $\mathbf{v}_d$. 
These 
processes can be captured using a transport equation: 
\begin{align}
\frac{\partial n}{\partial t} 
+ \underbrace{ (\boldsymbol{u} 
 + \langle \boldsymbol{v}_{d} \rangle) \cdot \nabla n}_{\text{advection \& drift}} 
- \underbrace{\boldsymbol{v}_{s} \cdot \nabla n}_{\text{streaming}} 
- \underbrace{\nabla \cdot \left( \boldsymbol{K} \cdot \nabla n \right)}_{\text{diffusion}} 
- \underbrace{\frac{\partial}{\partial \gamma} 
\left(b n \right)}_{\shortstack{\text{\footnotesize heating} \& \\ {\footnotesize \!\! cooling}}}
= \underbrace{Q - \Lambda n}_{\text{sources \& sinks}}. 
\label{eq:general_trans_eq}
\end{align} 

Here, $n = n(\gamma, \boldsymbol{x}, t)$ 
is the phase space density of CRs 
with energy $\gamma = E/m c^2$ at a time $t$ and position $\boldsymbol{x}$, where $m$ is the rest mass of the particle. CR particles are supplied by sources at a rate $Q$ and are removed by interactions at a rate $\Lambda$. 
The destruction term $\Lambda$ accounts for processes that 
fully remove particles from the population. 
This can occur either by a significant energy loss 
 in a single interaction 
 or by the complete destruction of the particle 
 (cf. the pion-production processes in Section~\ref{sec:hadronic_interctions}). 
The term $b$ is the total rate of continuous energy changes experienced by the CRs (cooling or heating), including adiabatic processes. 
This term also accounts for CR interactions that lead only to a small change in particle energy, 
  which are well approximated as a continuous heating or cooling process 
  (e.g., photo-pair Bethe--Heitler processes). 

\subsection{Neutral Particle Propagation}
\label{sec:neutral_prop}
 
Neutral particles 
do not experience strong deflections in magnetic fields. 
They 
  propagate ballistically 
  with a streaming velocity close to the speed of light, i.e., $|\boldsymbol{v}_s| \approx c$. 
  They can still engage in certain heating and cooling processes. 
CR neutrons, for instance, 
  can undergo some of the same heating and cooling processes as CR protons
  (e.g., in explosive environments 
  or during cosmological propagation). 
Similarly, they can be destroyed 
  by interactions 
  with baryons or radiation fields (see Section~\ref{sec:hadronic_interctions}),  
  and can 
  also be injected into astrophysical environments. 

The transport equation 
  for CR neutrons  
  may take the form
\begin{align}
\frac{\partial n}{\partial t} 
- \boldsymbol{v}_{s} \cdot \nabla n 
-\frac{\partial}{\partial \gamma} \left(b n \right) 
= Q - \Lambda n .
\label{eq:neutral_tran_eq}
\end{align} 

The life time of a neutron is insignificant  
  compared with cosmological timescales 
  (relevant for adiabatic losses 
   due to the expansion of the Universe)  
  and the dynamical timescales of their host environments 
  (where energy transport is often governed 
    by advection and diffusion processes). 
We therefore neglect cooling effects   
 for the transport of free neutrons 
 in our calculations (see also Section~\ref{sec:neutron_interactions}). As the prescription 
   for the transport process we adopt 
   is generic to neutral hadrons and nucleons,  
   any temporarily neutral baryon-number carrier streams ballistically and is not magnetically confined during its neutral phase. 

\subsection{CR Propagation with Charged--Neutral Switching}
\label{sec:general_prop}

In astrophysical structures 
 threaded by magnetic fields, CR protons diffuse 
 while CR neutrons propagate 
 ballistically and are practically unrestrained. 
Hadronic interactions   
  that temporarily convert protons to neutrons  
  therefore introduce  
  changes in the mode of CR propagation. 
  
Due to time dilation, 
  ultra-relativistic CR neutrons 
  can traverse 
   substantial distances 
  before they eventually decay 
  into charged protons and leptons. Depending on their 
  energy, they can traverse structures the size of galaxies, 
  clusters or even filaments. 
The characteristic widths of filaments 
and the sizes of clusters/super-clusters 
are usually of order a few Mpc. The crossing time  
  of ultra-relativistic neutrons  
  is therefore a few Myr. 
This is much shorter than 
  the dynamical and evolutionary timescales 
  of these systems,  which is typically of the order of a few 100~Myr. 
The disparity between the propagation timescales of CR neutrons and the evolutionary timescales of astrophysical structures allows simplifications to be made when 
including neutron–proton conversions in 
a CR propagation framework. 

Conversions arise from discrete interaction events. 
They can be treated as a separate stochastic term, 
and handled independently from deterministic transport processes. 
The resulting ballistic transport of neutrons manifests as stochastic jumps. 
To first approximation, CR transport with neutron–proton conversions can be described by a stochastic differential equation (SDE)\endnote{SDEs are widely used to model fluctuating systems, e.g., in renewable-power variability~(e.g., \cite{Anvari2016NJPh}), financial markets~(e.g., \cite{Zhou2021PhyA}), and biological systems~(e.g., \cite{Ghasemi2006physics}).} with Poisson jumps~\cite{sun2013mathematical, Movahed2024}. This approach allows 
the substantial numerical complexity of 
solving a two-species coupled system of equations 
to be bypassed without compromising the validity of our results. 

For the astrophysical situations relevant 
  to this study, 
  the transport SDE has two components:   
  one is deterministic, the other is stochastic. 
The two components are mutually 
  time exclusive 
  in the representation that CRs 
  are a single ensemble of charged protons 
  (neutrons not explicitly modeled here, apart from the propagation 
   jumps they introduce when in temporary ballistic motion). 
The deterministic component follows 
  from Equation~(\ref{eq:general_trans_eq}).  
The stochastic  
production of neutrons is 
   a Poisson jump in the SDE.  
The SDE may therefore be expressed as   
\begin{align}
 \frac{\partial n}{\partial t} = 
\underbrace{\left[ \nabla \cdot \left( \boldsymbol{K} \cdot \nabla n \right) + \boldsymbol{v}_s \cdot \nabla n 
 - (\boldsymbol{u} + \langle \boldsymbol{v}_d \rangle) \cdot \nabla n 
 + \frac{\partial}{\partial \gamma} \left(b n \right) + {Q} \right]}_{\text{deterministic}} 
+ \underbrace{h(\gamma, \boldsymbol{x}, t) \;\! \frac{\mathrm{d}J_t}{\mathrm{d}t}}_{\text{stochastic}} 
. 
\label{eq:sde_general}
\end{align}  

In this expression, 
  the differential term ${{\rm d}J_{t}}$ 
   takes values of 1 or 0, 
   with a probability of $\lambda {\rm d}t$ that it is 1 and $(1-\lambda {\rm d}t)$ that it is zero. 
When ${{\rm d}J_{t}}/{{\rm d}t}$ 
  is averaged over many jumps, 
it gives $\lambda$ as the characteristic rate 
   of the Poisson process. 
The parameter ${Q}$ 
   specifies the rate of (deterministic) 
   CR proton injection by astrophysical sources. 
It does not include secondary protons     
  that arise from neutron decay. 
The decay of neutrons 
  is not instantaneous, 
  and the stochastic production 
  of secondary protons are included 
  in the model as part of the jumps. 

The jump coefficient in the SDE 
  is given by  
\begin{align} 
  h(\gamma, \boldsymbol{x}, t) 
  = -n(\gamma, \boldsymbol{x}, t) \;\! 
   \Lambda \;\! \Delta t 
   + q(\gamma, \boldsymbol{x}, t) 
   \;\! \Delta t . 
\end{align}

The parameter $\Lambda$ 
  specifies the rate of hadronic interactions. 
It accounts for CR neutron production 
  and their subsequent decay to secondary protons. 
  Hadronic interactions can also produce 
  secondary CR protons directly,  
  but these 
  do not substantially contribute to the particle ensemble or affect our results 
  when included in our calculations\endnote{The 
secondary protons have energies lower than 
  the parent protons. High p$\gamma$ interaction rates 
  are only possible   at the highest of energies we consider. 
For primary CRs   with a broken power-law or 
  a power-law energy spectrum,   secondary protons 
  are always sub-dominant   to the primary CR protons. 
Secondary CR protons may, however,   be non-negligible    if the system is `optically thick' 
  to ultra-high-energy CR interactions   and the CRs have an unusually flat 
  spectrum,   such that there are substantial   amount of extremely high-energy 
  CR particles. In such a situation,  consideration 
  of the contribution by   directly formed secondary 
  CR protons would be required, 
  especially if modeling  the high-energy CR spectral component.}. 
The production of neutrons   in hadronic interactions 
  is specified by $q(\gamma, \boldsymbol{x}, t)$, 
  which gives the redistribution function 
  of the protons from the decay 
  of secondary neutrons. 
We express it as 
\begin{align}
    q(\gamma, \boldsymbol{x}, t) = \int_{\gamma_{\rm p}} 
     {\rm d}\gamma_{\rm p}
    \int_{\boldsymbol{x}} 
    {\rm d}\boldsymbol{x}   \ 
    g(\gamma; \gamma_{\rm p}) 
     \Lambda n(\gamma_{\rm p}, \boldsymbol{x}, t) \left\langle \delta \left( \boldsymbol{x} - \hat{\boldsymbol{x}} \gamma c \tau_{\rm neu} \right) \right\rangle_{\Omega}    ,
\end{align} 
where ${\boldsymbol{\hat x}} 
  = \boldsymbol{x}/|\boldsymbol{x}|$ 
   is the unit normal vector 
   of proton propagation, 
   and $\Omega$ is the deflection angle 
  of the secondary neutron 
  with respect to ${\boldsymbol{\hat x}}$.  
The angle-averaged delta-function $\delta$ 
  accounts for the addition of neutrons 
  from all directions 
  that will decay into a proton 
  at location $\boldsymbol{x}$. 
   $g(\gamma; \gamma_{\rm p})$ specifies 
  the fraction of secondary neutrons 
   of energy $\gamma$ produced by a primary CR proton of energy $\gamma_{\rm p}$ 
    in a hadronic interaction. 

Secondary neutrons 
  are converted back into protons 
  via decays, 
  and there is negligible energy loss 
  when the particles are in the neutron phase. 
We model the neutron decay using a Monte Carlo algorithm, 
   where the lifespan of a neutron is governed by 
   a stationary Poisson process, 
   i.e., sampled from an exponential distribution 
   in the particle co-moving frame, 
   with a mean lifetime of $\bar{\tau}_{\rm neu} = 879.6\pm0.8~{\rm s}$ \cite{PDG2024PR}. 

We model $g(\gamma; \gamma_{\rm p})$ 
  separately for pp and p$\gamma$ processes. 
For pp processes, 
we use {\tt Aafrag} 2.01~\cite{2023CoPhCKachelriess},  
which provides neutron production spectra derived from the pp interaction event generator {\tt QGSJET-II-04m}~\cite{Ostapchenko2011PhRvD, Ostapchenko2013EPJWC, Kachelriess2015ApJ}. For p$\gamma$ processes, 
  we use interpolated inclusive neutron production cross sections obtained 
  with the Monte Carlo photo-hadronic event generator, {\tt SOPHIA}~\cite{Mucke2000CoPhC}, which allowed neutron production rates to be pre-computed up to $E_{\rm max} \sim 2 \times 10^{19}\;\!{\rm eV}$. At higher energies, the neutron yield function was approximated by extrapolating its overall normalization and peak position from the highest-energy reliable {\tt SOPHIA} runs.  

The integration of Equation~(\ref{eq:sde_general}) gives  
\begin{align}
    n(\gamma, \boldsymbol{x}, t) 
    = & n_0(\gamma, \boldsymbol{x})
    + \int_0^t  {\rm d}t' \ 
     \nabla \cdot \left( \boldsymbol{K} 
       \cdot \nabla n \right)
    + \int_0^t   {\rm d}t'  \ 
      \boldsymbol{v}_{s} \cdot \nabla n    
    - \int_0^t {\rm d}t' \ 
     (\boldsymbol{u} + \langle \boldsymbol{v}_{d} \rangle)\;\! \cdot \;\! \nabla n 
    \nonumber \\ 
    & \hspace*{1cm}
    + \int_0^t {\rm d}t' \ 
     \frac{\partial}{\partial \gamma} \left(b n \right) 
    + \int_0^t {\rm d}t' \  Q  
    + \int_0^t  {\rm d}J_{t'} \ 
     h(\gamma, \boldsymbol{x}, t')  .  
    \label{eq:integral_eq} 
\end{align}

Initially, no particles are present, i.e.,  
  $n_0(\gamma, \boldsymbol{x}) = n(\gamma, \boldsymbol{x}, t=0) = 0$, 
  and CRs are injected solely as charged particles 
  by sources through the source term $Q$. 
 
Stochastic population equations with jumps  
  (which include the SDE constructed for this work) 
  generally do not have 
  explicit closed-form solutions 
  \cite{sun2013mathematical}. 
Conceptually, the integral equation 
  (Equation~(\ref{eq:integral_eq})) can be solved numerically using a discrete implicit Euler-type 
  scheme 
  to find an approximate solution: 
\begin{align}
N^{k+1}_t = N^k_t + \left[ \nabla \cdot \left( \boldsymbol{K} \cdot \nabla N^{k+1}_t \right)
+ \boldsymbol{v}_s \cdot \nabla N^{k+1}_t 
- (\boldsymbol{u} + \langle \boldsymbol{v}_d \rangle) \cdot \nabla N^{k+1}_t 
\right.  \nonumber \\ 
 \left.  + 
 \frac{\partial}{\partial \gamma} \left(b N^{k+1}_t \right)
  + {Q}^k_t \right] \Delta t + h(t, N^k_t) \Delta P_n \ ,
  \label{eq:integral_eq_stoch}
\end{align} 
with initial value $N^{0}_t = n(\gamma, \boldsymbol{x}, t=0) = 0$.  Here, $N^k_t$ is the approximation to 
$n(\gamma, \boldsymbol{x}, t_k)$ for $t_k = k\Delta t$, where $ \Delta t$ is the time increment. 
The term $\Delta P_n = P(t_{k+1}) - P(t_k)$ denotes the increments of the Poisson process. 
In this work, we solve the deterministic terms 
  using the stiff Ordinary Differential Equation 
  (ODE) solver {\tt RADAU5}~\cite{hairer2010solving},  
  with adaptive internal time steps.  
We apply the jump term explicitly 
  at the end of each global time step, 
  where the jump $h(\gamma, \boldsymbol{x}, t) \Delta P_n$ is implemented 
  via Poisson sampling over effective sub-volumes of each cell of the discretized computational domain. 
A detailed description 
  of the computational scheme 
  developed to solve the stiff SDE 
  is provided 
  in Appendix~\ref{sec:appendix_a}.

\section{CR Confinement and Escape from Magnetized Structures}
\label{sec:sec3}

\subsection{Illustrative Applications}
\label{sec:demo_applications}

We illustrate the confinement 
  and escape of CRs 
  using three hierarchical structures: 
  cosmological filaments, galaxy clusters and 
  galaxies. For galaxies, we consider the escape of CRs 
  in both normal galaxies (of the main sequence, Milky Way-like configuration), 
  and starburst galaxies. This is to highlight the impact of the 
  variation of the properties of different galaxy types on CR confinement, escape and calorimetry. 
These structures span 
    from scales of several Mpc down to kpc.  
CRs produced in galaxies 
  can seep into the surrounding halos, and then into 
  the clusters and filaments in which they reside. 
   Understanding these CR exchanges between structures is essential for building complete models of these 
   environments and for assessing the dynamical role CRs may play within them. 
   
Current formulations of CR transport 
  generally put focus on the local effects 
  in particular systems: 
  the interstellar media of Milky Way 
  and Milky Way-like galaxies 
  (e.g., \cite{Ponnada2024MNRAS, Amato2018AdSpR}), starburst galaxies~(e.g., \cite{Krumholz2020MNRAS, Dorner2023A&A}), 
  the circum-galactic medium (CGM)  (e.g., \cite{Butsky2023MNRAS, Roy2025arXiv251021699R}), 
  galaxy clusters (e.g., \cite{Ruszkowski2023A&ARv, Ensslin2011A&A}), in particular their feedback via active galactic nuclei (AGN) jets and bubbles~\cite{Yang2019ApJ, Lin2023MNRAS}, holistic propagation through multi-component media~(e.g., \cite{Kempski2023MNRAS, Butsky2024MNRAS, Ewart2025arXiv250719044E}) or specific substructures within galaxies~(e.g., \cite{Owen2021ApJ, Pineda2024A&A}). 
One of the technical challenges 
  in these studies 
  is to obtain 
  a reliable treatment of 
  the CR transport process 
  when magnetic fields 
  have structures 
  over a wide range of spatial scales. 

Galaxies, galaxy clusters and filaments 
  are hierarchical structures 
  within a larger ecosystem of the cosmic web. 
They have distinctive  magnetic properties in field strength and also in topology.  
A multi-scale treatment 
  of CR transport 
  across these structures 
  with the inclusion of exchanges  
  across their interfaces 
 requires a formulation  
  appropriate for system parameters
  over large dynamical scales.  
Taking the ecosystem 
  of galaxies and their CGM 
  (e.g., \cite{Owen2023Galax})
  as an example, 
  CRs are produced by star formation, supernova remnants, and AGN. 
They are then transported by advection in galactic outflows and by diffusion along and across magnetic fields that thread and cocoon galaxies and interconnect neighboring systems. 
The relevant physics 
  must be consistently incorporated  
  in the formulation, 
  in particular those that 
  determine the CR exchange 
  between galaxies, 
  galactic halos, 
  and \mbox{their environments.} 

Two steps upward in the scale hierarchy  
  are cluster--filament systems.  
Recent observations 
 have begun to reveal magnetic fields
in cosmological filaments 
 with some topological information 
 (in addition to field strength estimation;~(e.g., \cite{Vernstrom2021MNRAS, Carretti2022MNRAS, Vernstrom2023SciA})). New 
 techniques~(e.g., with synchrotron intensity gradients \cite{Hu2024NatCo}) and improved data have also revealed rich field structures in galaxy clusters, including 
locally strong, ordered field layers superposed on turbulent background components~\cite{Anderson2021PASA, Rajpurohit2022A&A, Balboni2023A&A} with substantial cluster-to-cluster diversity that likely reflects the dynamical histories of individual systems~(e.g., \cite{Liou2025arXiv251026218L}). Despite these advances, a 
systematic theoretical investigation of CR transport across these environments is still lacking.

 Comparing the magnetic fields in clusters and filaments highlights several subtle issues at cluster--filament interfaces 
  (e.g., \cite{Wu2024Univ, Owen2025arXiv250415802O}). In particular, these interfaces are not simple scaled-up analogues of galaxy-CGM or galaxy–group boundaries. CR-transport 
scenarios and formulations developed for galaxies therefore cannot be directly translated to the cluster--filament regime. Before  
 we can model CR transport and calorimetry on these scales~\cite{Wu2024Univ}, 
 we must first establish how CRs are exchanged across 
 cluster--filament interfaces. 
 This CR transfer sets the input to filaments and voids, which act as reservoirs of fossil CRs that record the non-thermal power generation history of the Universe.
 
As a first study, 
we investigate the transport of the most energetic CRs on cosmological scales. We focus on neutron-mediated escape as a mechanism to break confinement and enable CR transfer between different partitions of the cosmological ecosystem. We consider an energy range between $10^{12}$ eV and $10^{21}$ eV, covering the lowest energies at which CRs are expected to build-up in components of the large-scale structure of the Universe~\cite{Wu2024Univ} to energies that bracket the most energetic CRs ever detected~\cite{Bird1995ApJ, TA2023Sci}. 
Our emphasis is on CR exchange across structural boundaries, rather than on the detailed CR microphysics within galactic scales where galactic-scale bulk flows (winds and accretion) and their feedback dynamics would need to be modeled explicitly. 
We also 
ignore long-distance advective transport, such as that by giant radio jets launched by supermassive black holes, which can extend beyond Mpc scales and even ``puncture'' the cosmic web~\cite{Oei2024Natur}. 
  To avoid additional complexity that would obscure the essential physics and signatures of CR transport between galaxies, clusters, filaments and their interfaces, we also do not include the back-reaction of CRs on the magnetic fields or on the dynamical evolution of these structures.

We adopt a generic geometrical representation, 
  where 
  galaxies are modeled as flattened cylindrical disks 
  (pancake-like),  
  clusters are spheres, 
  and filaments are elongated cylinders (tube-like).  
Their radiation energy densities take representative values listed in Table~\ref{tab1:physical_params}. The gas density is treated as radially uniform throughout each of the structures considered, and is fixed to the values given in Table~\ref{tab1:physical_params}. The vertical gas distribution is uniform in the case of filaments but decays exponentially with height in our two galaxy models, with the galaxy height in Table~\ref{tab1:physical_params} being used as a scale height for the exponential profile. Each structure hosts a magnetic field with strength spatially varying with the density profile, and set to central values listed in Table~\ref{tab1:physical_params}. Primary CR protons are injected isotropically by sources located within the magnetized volume, defined for of each structure (galaxy, cluster, or filament; cf. Table~\ref{tab:geometrical_params}). Their spatial distribution follows the gas density profile. The geometry, diffusion properties, and the definitions of the escape surfaces used to determine particle escape are summarized in Table~\ref{tab:geometrical_params}.  

\begin{table}[H] 
\caption{Summary of fiducial parameters adopted for each of the astrophysical and cosmological structures considered in our calculations, chosen to roughly characterize each environment. For CR interactions via p$\gamma$ processes with radiation fields, we include starlight and dust-reprocessed components associated with the extragalactic background light or interstellar radiation fields appropriate to each structure. In addition, we account for CR interactions with the cosmic microwave background (CMB), using its properties at a redshift, $z$, specified for each of our results.}  

	\begin{adjustwidth}{-\extralength}{0cm}

		\begin{tabularx}{\fulllength}{LCCCC}

			\toprule
\multirow[m]{2}{*}{\textbf{Environment}}	& \multicolumn{2}{c}{\textbf{Radiation Energy Density \textbf{[\boldmath{${\rm eV\;\!cm}^{-3}$}]}}} &  \textbf{Gas Density} & \textbf{Magnetic Field} \boldmath{$^{\ddagger}$} \\
& \textbf{Starlight} & \textbf{Dust} & \textbf{n} \textbf{[cm\boldmath{}$^{-3}$]} & \boldmath{$\langle |\mathbf{B}| \rangle$} \textbf{[\boldmath{$\upmu$}G]} \\
\midrule

Normal Galaxy $^{\star}$   & 0.67   & 0.31   & $1.0$       & $10.0$ \\
Starburst Galaxy $^{\star \star}$   & 67.0   & 31.0   & $300.0$       & $500.0$ \\
Cluster $^{\dagger}$  & 1.0 & 0.5  & $0.003$   & $5.0$  \\
Filament $^{\dagger}$  & 0.3   & 0.14   & $0.0001$   & $0.05$ \\
	\bottomrule
		\end{tabularx}

	\end{adjustwidth}
\label{tab1:physical_params}

\noindent{\footnotesize{$^{\star}$ Radiation-field energy densities are scaled from extragalactic background light (EBL) values so that the total energy density is close to that of a Milky Way-like galaxy. The radiation fields are modeled as modified blackbodies with characteristic temperatures of $7100\,{\rm K}$ (starlight) and $62\,{\rm K}$ (dust), chosen to match the dominant components of the EBL. For filaments, the radiation-field energy densities follow the combined stellar and dust contributions to the EBL in Ref.~\cite{Dermer2009herb}. $^{\star \star}$ Radiation fields and density scaled by 100 compared to the Normal Galaxy model. This approximately captures the conditions of nearby starbursts~(e.g., \cite{YoastHull2016MNRAS}). Magnetic field strength is an order of magnitude larger than that of a normal galaxy, informed by approximate magnetic field strengths reported for nearby starbursts---e.g., M82~\cite{Persic2025A&A}. $^{\dagger}$ In clusters and filaments, photon fields of starlight and dusts  
are strongly centrally peaked and become subdominant to the CMB outside the inner regions. We therefore treat our adopted energy densities as a conservative, spatially averaged target field appropriate to clusters at intermediate radii and filaments in their spines~(see \cite{Pinzke2011PhRvD}). $^{\ddagger}$ Magnetic field strengths are chosen to be representative of spiral galaxies~(e.g., \cite{Beck2015A&ARv}), 
intra-cluster fields~(e.g., \cite{Carilli2002ARA&A,Govoni2004IJMPD,Osinga2022A&A}), and cosmic filaments~(e.g., \cite{Ryu2008Sci, Vacca2018Galax}).}}

\end{table}

This treatment considers a simplified scenario. 
The magnetic field in real cosmic filaments, clusters, and galaxies 
has multi-scale structures, 
including ordered components, 
domains with spatially varying 
 coherence lengths, 
and turbulence-induced incoherent fluctuations.   
The relative strengths 
of the ordered and disordered field components  
(such as those due to turbulence) 
affect the effective 
diffusion tensor (in particular, the ratio between parallel and perpendicular diffusion relative to the large-scale field vector) 
and 
therefore the escape probability of the 
charged particles.   
The precise onset energy and 
magnitude of the escape fractions derived here should be 
considered model-dependent. Our aim is not to provide a fully turbulence-resolved transport calculation 
for each environment. 
Instead, we aim to isolate the qualitative effect of temporary charged--neutral switching. 
The exact quantitative importance of this channel will depend on the magnetic geometry and turbulence 
level, and will likely vary between specific environments. This could especially be the case in cosmological filaments 
where there may be large-scale field ordering~\cite{Carretti2022MNRAS}.

\begin{table}[H]
\caption{Geometrical configurations and CR diffusion parameters adopted for the three fiducial models. 
Galaxy parameters are chosen to approximate a Milky Way-like system. 
Characteristic filament lengths and widths are guided by typical mean values of their central spine regions from 
simulations (e.g., Ref.~\cite{GalarragaEspinosa2024A&A}), with cluster sizes comparable to filament widths~\cite{Kravtsov2012ARA&A}. In the case of galaxies, CRs are injected uniformly throughout the volume of the structure. In the case of clusters, CRs are injected up to half of the cluster radius to reflect the clustering of CR sources and the dominance of a potential strong source associated with a bright central galaxy. For filaments, our model captures the central spine region where sources would be concentrated. Therefore we inject the CRs uniformly out to 90\% of the filament radius, and uniformly along the length of the structure.}  
   	\begin{adjustwidth}{-\extralength}{0cm}
		\begin{tabularx}{\fulllength}{LLm{5cm}<{\centering}CC}
			\toprule
       \textbf{Environment}  & \textbf{Geometry} & \textbf{Dimensions} & \boldmath{$|\boldsymbol{K}_{||}|/|\boldsymbol{K}_{\perp}|$} $^{\textit{\textbf{(a)}}}$ & \boldmath{$|\boldsymbol{K}|^{\rm TeV}$}  
       $^{\textit{\textbf{(b)}}}$ 
   \\    &  & & &  \textbf{[cm\boldmath{$^2$} s\boldmath{$^{-1}$}]}    \\
       \midrule

        Galaxy & Cylindrical & $R_{\rm gal} = 15.0$ kpc; \;\! $H_{\rm gal} = 4.0$ kpc & 1 & $1\times 10^{30}$ \\
        Starburst & Cylindrical & $R_{\rm gal} = 15.0$ kpc; \;\! $H_{\rm gal} = 2.0$ kpc & 1 & $3\times 10^{28}$ \\
        Cluster & Spherical & $R_{\rm cl} = 2.0$ Mpc & 1 & $3\times 10^{30}$ \\
        Filament & Cylindrical & $R_{\rm fil} = 0.5$ Mpc; \;\! $\ell_{\rm fil} = 20.0$ Mpc & 10 & $3\times 10^{30}$ \\

        	\bottomrule
		\end{tabularx}
	\end{adjustwidth}
	\noindent{\footnotesize{\textbf{Notes:} $^{\textit{(a)}}$ \textbf{Diffusion anisotropy.} 
For galaxies and clusters, we invoke isotropic diffusion to clearly isolate the neutron-mediated escape channel. However, our results in these cases are relatively
insensitive to the exact configuration of their magnetic fields for any sensible choice of diffusion anisotropy.
Cosmological filaments
are extended structures, and the large observed polarization fractions from filament radio and microwave emission~\cite{Wolleben2021AJ,Bennett2013ApJS} indicate they are
threaded by highly ordered magnetic fields oriented roughly along their spine~\cite{Carretti2025Univ}. This may affect CR propagation on large scales. 
Filaments likely host super-Alfv\'{e}nic turbulence on large scales where
$B \sim \mathrm{nG}$ (with Alfv\'{e}nic Mach number $M_{\rm A} \sim 10$). In more 
magnetized filaments (e.g., near cluster interfaces) where
$B \sim$ 20$-$60$\,\mathrm{nG}$~\cite{Carretti2022MNRAS}, the turbulence may be
trans-Alfv\'{e}nic, but the cascade should still become sub-Alfv\'{e}nic below a transition
scale. We therefore consider that CR scattering on small scales ($\ll R_{\rm fil}$) can be
described using sub-Alfv\'{e}nic MHD turbulence scalings~\cite{Lazarian2023FrASS}. In this
regime, the perpendicular diffusion coefficient is suppressed by a factor
$M_{\rm A}^4$~\cite{Yan2008ApJ} such that
$|\boldsymbol{K}_{\parallel}|/|\boldsymbol{K}_{\perp}| \sim 10$ for modest
Alfv\'{e}nic Mach numbers. We adopt this value as a representative CR diffusion anisotropy
along the filament. A broader exploration of $|\boldsymbol{K}_{\parallel}|/|\boldsymbol{K}_{\perp}|$, 
coherence length, and turbulence level is deferred to future work. The filament anisotropy adopted here should be viewed as a representative fiducial choice rather than a unique prediction $^{\textit{(b)}}$ \textbf{Normalization of the diffusion coefficient (at 1 TeV).}
A typical value for the Milky Way is adopted for the Normal Galaxy model, while a suppressed 
value more typical of star-forming regions is adopted for the Starburst Galaxy model~(e.g.,~\cite{Semenov2021ApJ}). Faster diffusion with higher
coefficients is considered in large-scale
structures such as galaxy clusters~\cite{Wiener2019MNRAS,Habegger2025arXiv}, which is
reflected by our fiducial parameter choices. In our treatment, the magnitude of the CR diffusion coefficient is allowed to vary 
with CR energy, with a dependence 
$\propto r_{\rm L}^{1/2}$ (or $\propto E^{1/2}$), appropriate for diffusion in
Kraichnan-like compressible turbulence (e.g., \cite{Gabici2019IJMPD}). }}
\label{tab:geometrical_params}
\end{table}

\subsection{Escape Fraction and Multiplicities}
\label{sec:escape_criteria}

CR particles leave the system 
  by crossing a defined 
  ``escape surface'', 
  which is set by the 
  geometrical dimensions of each structure 
  (see Table~\ref{tab:geometrical_params}). 
To obtain a CR escape fraction 
  $f_{\rm esc}(E)$, 
  we tally the number of CRs that escape as a fraction of the total number injected. 
Escape can occur by spatial diffusion, ballistic streaming at ultra-high energies where scattering becomes negligible, or by conversion to neutrons that then free-stream out of the domain. 
  Advection is not included in our calculations. 
We treat these escape mechanisms as separate channels.

At each global timestep, we compute the number of CR particles that have left the structure across the chosen escape surface by comparing 
the total number of particles within the domain before and after the deterministic transport update. For cylindrical structures, we additionally decompose the escape into contributions from the sides and from the top and bottom caps. We use the discretized deterministic transport operator to estimate the 
outward flux through each portion of the boundary, and obtain 
the relative contributions of side and cap escape to the total deterministic losses. Streaming escape for charged particles is treated as the ballistic limit of this transport, when the effective propagation speed approaches the speed of light, $c$. 

Neutron escape is treated stochastically 
(i.e., the neutral-nucleon escape channel). Neutrons produced in hadronic interactions are sampled following a Poisson distribution   
  (see Appendix~\ref{sec:app_a_stochastic} for details). Each interaction converts a proton into one or more neutrons, to which we assign isotropic propagation directions in the center-of-momentum frame of the interaction. For each neutron, we draw a decay time from the exponential decay law. 
This sets the (straight–line) displacement 
  in the neutron phase 
  for a given neutron energy. 
If a neutron 
  crosses the escape surface before decaying, the subsequent neutron–proton conversion occurs outside the structure and the particle is counted as escaped, contributing to an energy-binned tally 
  $N_{\rm esc}^{\rm n}(E)$. If the neutron decays inside the structure, or if a neutron produced outside re-enters before decaying, we treat its decay products (protons) as secondary CRs and re-inject them into the grid rather than counting them as escaped. As with the deterministic component, 
  the surface of escape is recorded in the cylindrical geometry to distinguish whether particles escape via the sides or the top/bottom caps. 

We define the escape fraction 
  as the fraction of injected particles that eventually leave the system by any route. 
A given primary CR can either diffuse out, 
  escape by ballistic streaming as a charged particle, 
  or undergo at least one interaction that produces an escaping neutron which carries its baryon number outside.  
This gives an escape fraction: 
\begin{align}
f_{\rm esc}(E) =
\frac{
  N_{\rm esc}^{\rm diff}(E) +  N_{\rm esc}^{\rm str}(E)
  + N_{\rm esc}^{n}(E)
}{
  N_{\rm inj}(E)
} \,
\end{align}
where $E$ is the energy of the CR primary proton, $N_{\rm esc}^{\rm diff}$ is the number of CRs that escape by diffusion, 
$N_{\rm esc}^{\rm str}$ is the number that 
escape via streaming as charged particles (without undergoing a hadronic interaction), $N_{\rm esc}^{n}$ is the number that escape via interactions that produce neutrons, and $N_{\rm inj}$ is the number of CRs injected into the system, including secondary charged CRs that are injected by the decay of neutrons. 
To make the {neutron escape effect} explicit, we define {an escape factor specifically for the neutron channel,} to quantify the number of CRs that are released from a structure {by neutron-mediated jumps}: 
\begin{align}
A(E) =
\frac{ N_{\rm esc}^{n}(E)}{N_{\rm inj}^{\rm pri}(E)
}, 
\end{align}
which gives the mean number of neutron-channel escapes compared to injected primary CRs, $N_{\rm inj}^{\rm pri}$, at an energy $E$. When $A(E) \ll 1$, neutron-mediated release is inefficient and the structure effectively confines CR baryons. When $A(E) \approx 1$, each interacting CR proton that is lost from within the structure is   
replaced by baryons at the same energy escaping into high-order structures. 

To quantify the anisotropy of the escaping CR flux in cylindrical geometries, we define geometric fractions $X_{\rm side}(E) = { N_{\rm esc, side}^{\rm all}(E)}/{N_{\rm esc}^{\rm all}(E)
}$ and $X_{\rm end}(E) = { N_{\rm esc, end}^{\rm all}(E)}/{N_{\rm esc}^{\rm all}(E)}$, where 
$N_{\rm esc, side}^{\rm all}$ and $N_{\rm esc, end}^{\rm all}$ 
are the total numbers of escaping CR nucleons (primaries plus all secondaries) through the cylindrical side and end-caps, respectively. By construction, 
$X_{\rm side}(E) + X_{\rm end}(E) = 1$.  
The global primary-level escape fraction $f_{\rm esc}(E)$ is then decomposed geometrically as
$f_{\rm esc, side}(E) = f_{\rm esc}(E) X_{\rm side}(E)$ and 
$f_{\rm esc, end}(E) = f_{\rm esc}(E) X_{\rm end}(E)$. 
In this definition, {$f_{\rm esc, side}$ and $f_{\rm esc, end}$ 
give the probability that an injected primary ultimately contributes to a side- or end-escaping CR via any sequence of interactions and \mbox{secondary production}}. 

\subsection{Results}
\label{sec:results}

\begin{figure}[H]
\begin{adjustwidth}{-\extralength}{0cm}
\centering
{\includegraphics[width=9.0cm]{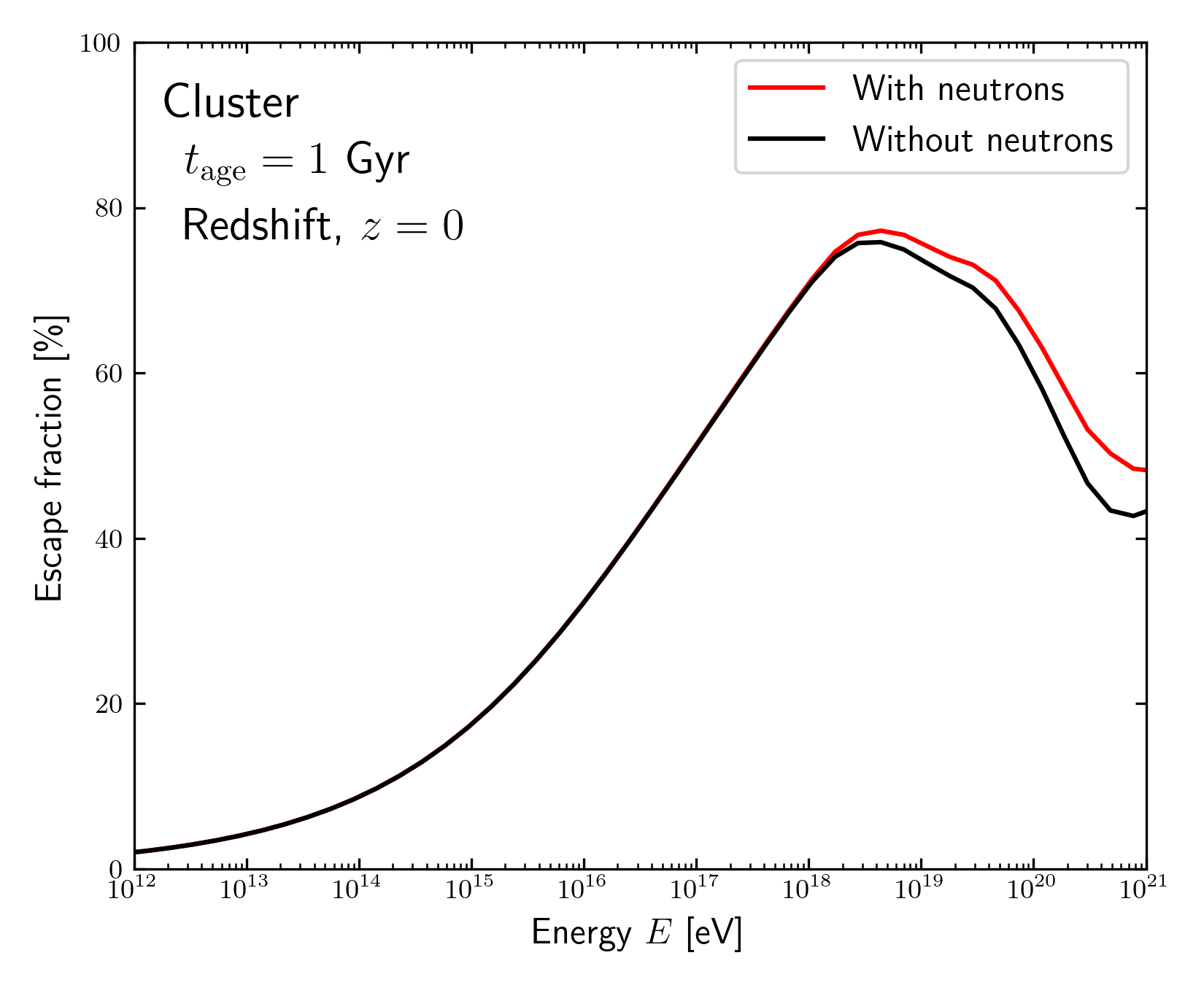}}
{\includegraphics[width=9.0cm]{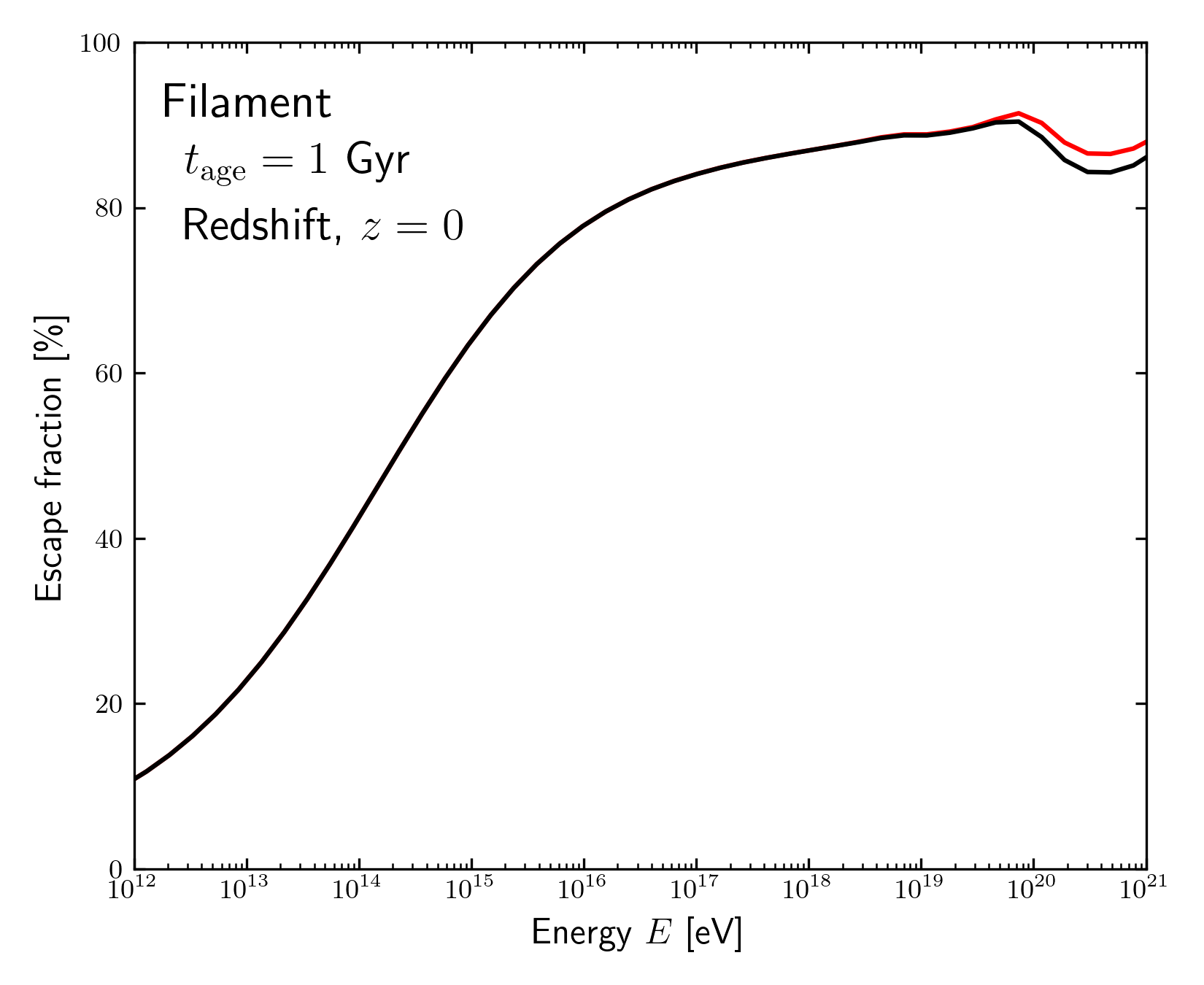}}
{\includegraphics[width=9.0cm]{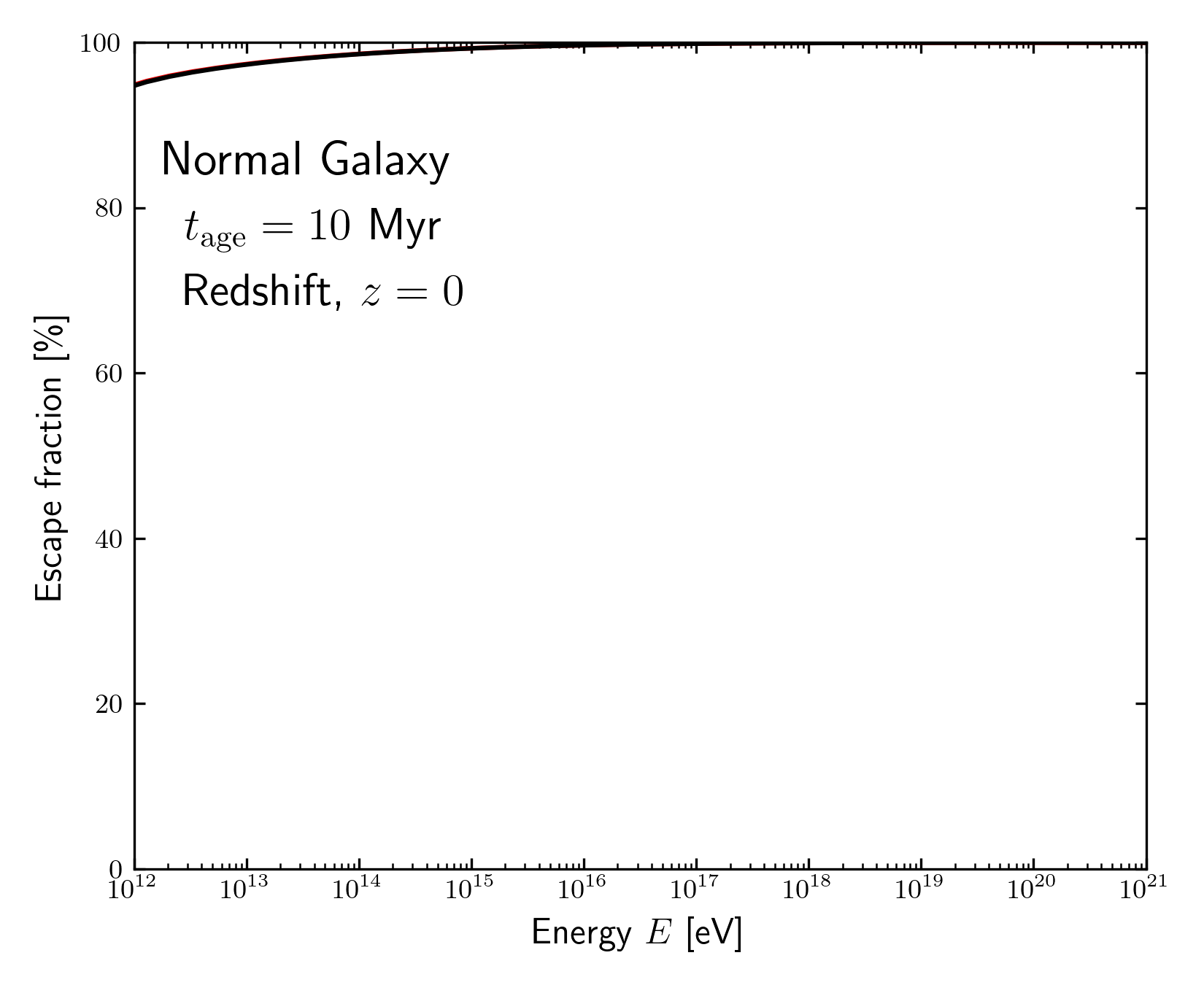}}
{\includegraphics[width=9.0cm]{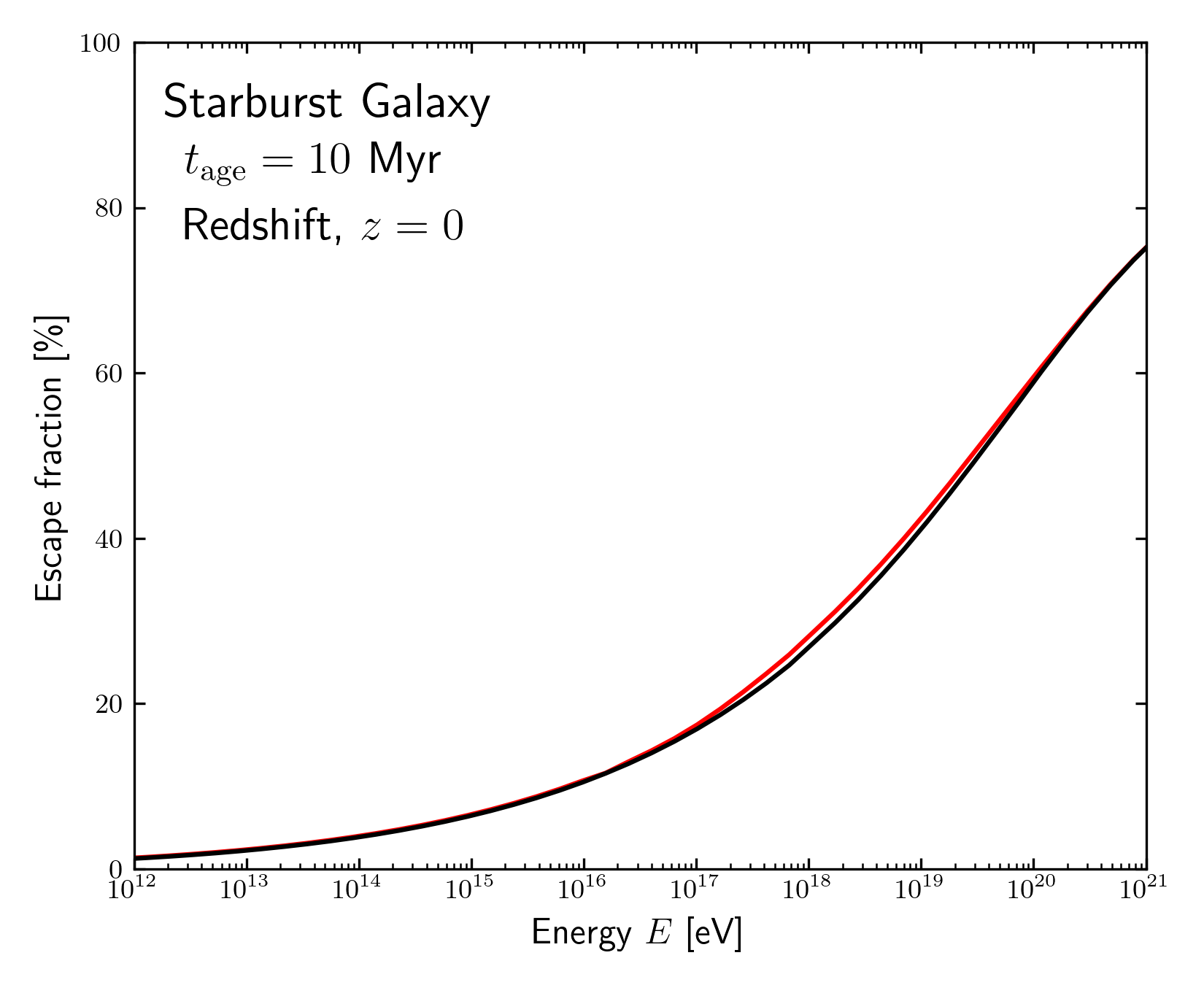}}
\end{adjustwidth}
\caption{Energy-dependent escape fraction, $f_{\rm esc}(E)$, of injected CR protons from the four characteristic environments considered here: a galaxy cluster, a cosmological filament, a ``normal'' (Milky Way-like) galaxy, and a starburst galaxy. The black curves show the standard treatment in which hadronic (pp and p$\gamma$) interactions act only as sinks for the charged CR proton population (``without neutrons''). The red curves include charged--neutral switching: neutrons produced stochastically in hadronic interactions propagate ballistically and are counted as escaped if they cross the escape boundary before decaying (``with neutrons''). The escape fraction includes contributions from diffusive and ballistic escape of charged particles and, when active, neutron-mediated escape. Neutron production has a marked impact on high-energy CR escape in all four systems. In clusters and filaments, the neutron-mediated escape channel is dominated by p$\gamma$ interactions. In galaxies, particularly starbursts (where magnetic fields are stronger and interstellar gas density is typically higher), pp interactions drive neutron production at lower energies, but they decay before escaping. Thus the escape channel is only marginal at intermediate energies. In clusters, neutron production 
causes a 
modest increase in escape relative to the no-neutron case between $\sim$10$^{18}$ eV under present \mbox{($z=0$) conditions.}}
\label{fig:fesc_all_1}

\end{figure}

Figure~\ref{fig:fesc_all_1} shows  
the energy-dependent escape fraction, $f_{\rm esc}(E)$, for starburst and normal galaxies, galaxy clusters and cosmological filaments, obtained from our calculations. 
We compare a ``standard'' 
treatment (where 
hadronic interactions act only to attenuate the confined charged CR populations) with a scenario where 
 interactions are allowed to produce secondary CR neutrons that propagate ballistically and may leave the system before decaying. We 
 evaluate the escape fractions at fiducial ages $t_{\rm age}$ of 10 Myr for the galactic models, 
 and 1 Gyr for the cluster and filament models. 
 These time snapshots 
 are chosen as representative intervals  
 over which the evolution of each system 
 is sufficiently insignificant,   
 while avoiding the unrealistic requirement that the calculations be  
 followed until even the slowest diffusing, most strongly confined low-energy CRs have fully converged on an escape fraction. 
 Over these timescales, the galactic 
 models have nearly approached a quasi-steady escape spectrum, 
 whereas the larger-scale structures in our calculations do not do so at low energies. 
 In the cases of clusters and filaments, 
 the relevant escape timescale can exceed a Hubble time, implying that the low-energy CR population in such systems would not in general be expected to have relaxed to a steady state in the present Universe~\cite{Pinzke2010MNRAS, Brunetti2014IJMPD, Wu2024Univ}. 
 However, at the high energies most relevant to the neutron-mediated channel, 
the simulation time exceeds both the longest relevant transport timescale in the confined phase 
(set by the slowest diffusing energies) and the characteristic crossing time of the structure ultra-relativistic particles propagating at the speed of light, $c$. Thus, 
 at the adopted snapshot, 
 all systems have evolved for long enough that their energy-binned escape tallies have converged 
 in the regime where neutron-mediated escape is important. 
 In this energy range, our results are therefore insensitive 
 to the precise end-time of the calculation because further evolution increases both 
$N_{\rm inj}(E)$ and the cumulative escaped counts proportionally. This makes the resulting 
 escape fractions useful for quantifying inter-structure CR transfer, 
 as they represent the long-term averaged leakage from a persistent  
 population of sources embedded within magnetized structures at a given epoch.  

 Across the environments considered, 
 the main qualitative trend is that neutron production 
 has little effect on CR escape from galactic-scale structures but can substantially 
 alter confinement in larger-scale systems. 
 The amount of escape enabled by the neutron channel, and the energy 
of its 
onset, depends on two main factors. These are 
(i) the probability that a parent CR proton interacts within the volume of the confining structure, and 
(ii) the probability that a neutron produced in an 
interaction traverses the distance to an escape surface before undergoing $\beta$-decay. 
 In starburst galaxies, 
 the probability of an interaction through the pp-channel is high, but this process becomes important at  
 energies where the resulting neutrons still cannot traverse an escape distance prior to decay. 
By contrast, in large-scale structures, neutron-mediated escape 
  partially decouples the fate of injected high-energy protons 
from magnetic confinement and strong photo-attenuation. 
In energy ranges where the standard treatment strongly suppresses CR escape due to interactions occurring faster than charged-particle escape, the neutron channel restores part of the escaping component 
by temporarily converting confined protons into free-streaming nucleons. 

\begin{figure}[H]
\begin{adjustwidth}{-\extralength}{0cm}
\centering
{\includegraphics[width=9.0cm]{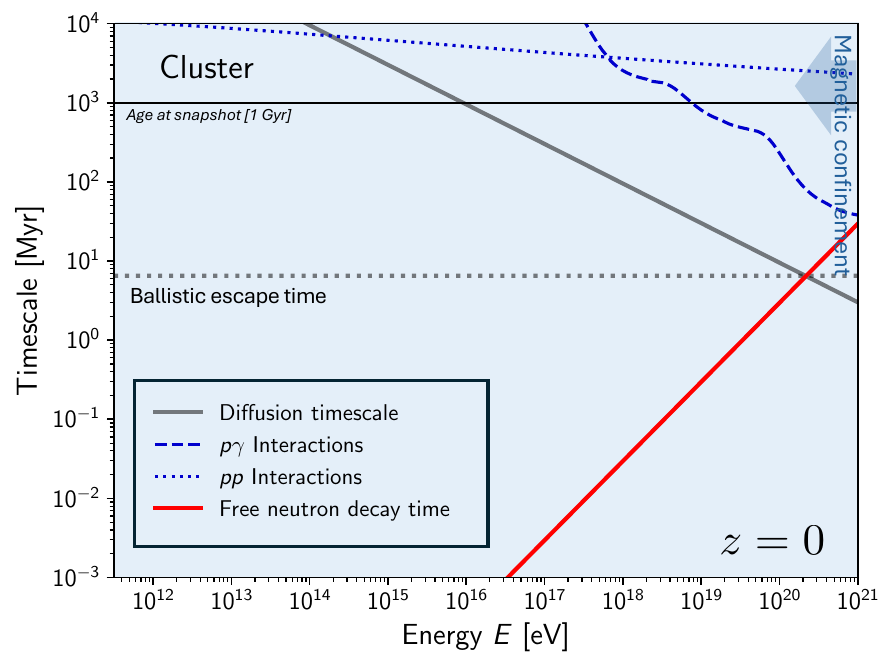}}
{\includegraphics[width=9.0cm]{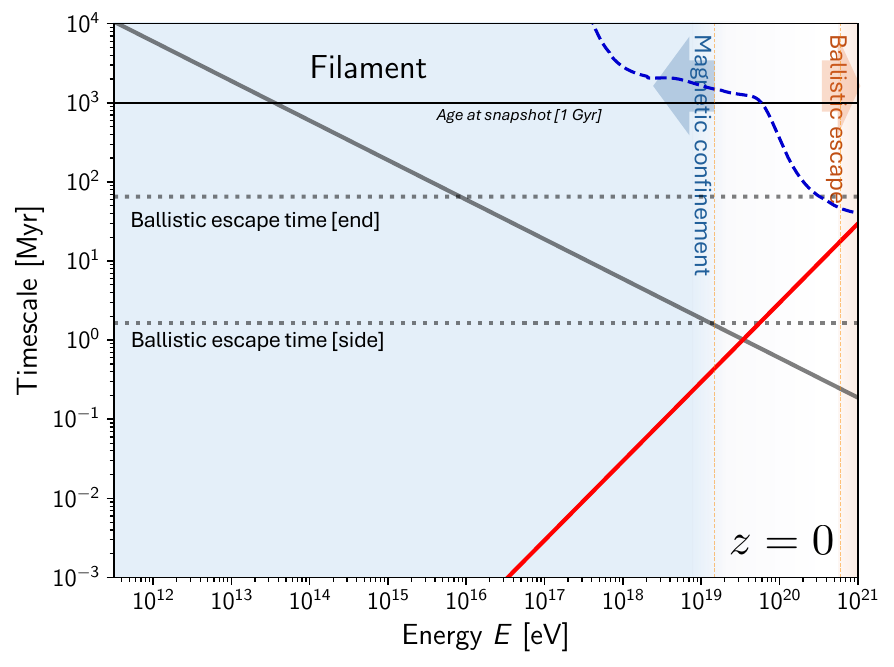}}\\
{\includegraphics[width=9.0cm]{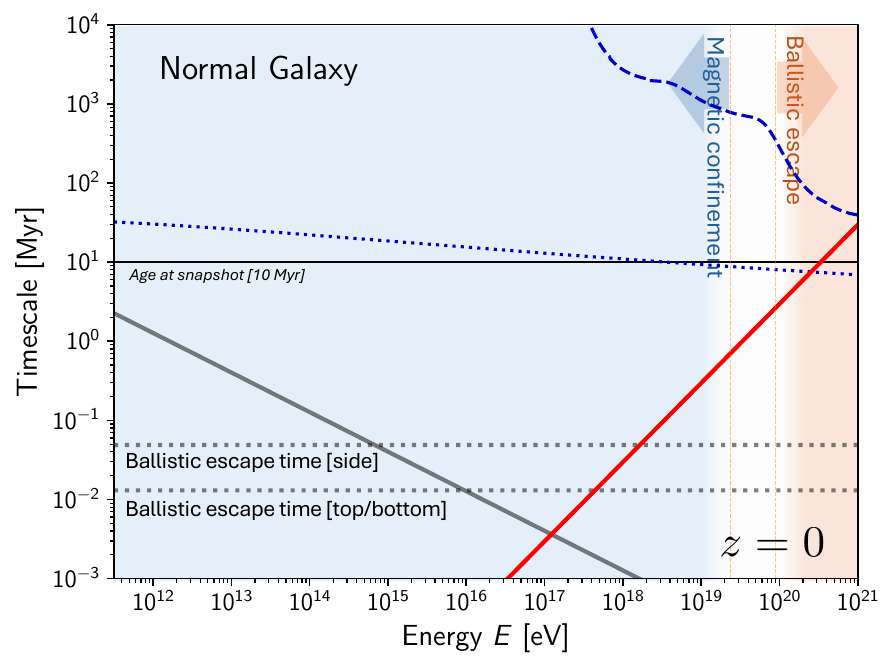}}
{\includegraphics[width=9.0cm]{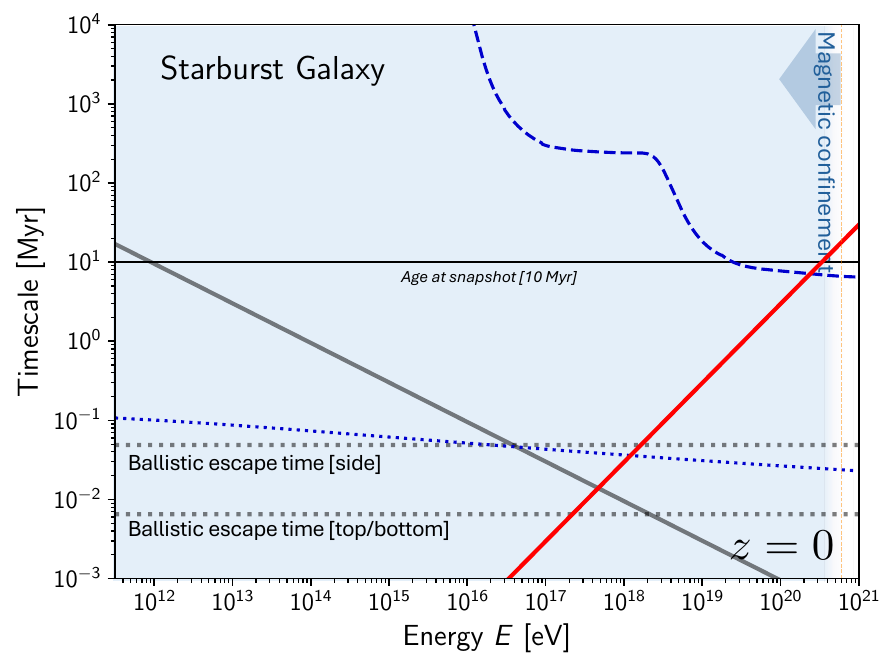}}
\end{adjustwidth}
\caption{Characteristic timescales governing CR confinement, hadronic interactions, and neutron-mediated escape in the four 
representative environments considered: a galaxy cluster, a cosmological filament, a normal (Milky Way-like) galaxy, and a starburst galaxy. 
In each panel we compare the charged-particle transport timescale, the pp and p$\gamma$ interaction timescales, 
the Lorentz-dilated free-neutron $\beta$-decay time, the ballistic crossing time to the relevant escape surface(s), and the fiducial snapshot age adopted for the escape calculations in Figure~\ref{fig:fesc_all_1}. For the cylindrical systems, 
separate ballistic crossing times are shown for escape through the side boundary and 
through the end-caps/top--bottom surfaces. The shaded vertical bands indicate the approximate transition from magnetically confined charged-particle transport to the ballistic regime, based on the particle energy }  
\label{fig:timescale_fig_1}

\end{figure}

 The results 
   {we have obtained} 
 can be more clearly understood with reference to the relevant timescales mediating escape,  
 shown in Figure~\ref{fig:timescale_fig_1}, where we compare the characteristic charged-particle transport 
 timescale, the pp and p$\gamma$ interaction timescales, the Lorentz-dilated free-neutron decay time, and the ballistic crossing time to the relevant escape surface, together with the snapshot age adopted in Figure~\ref{fig:fesc_all_1}. 
 The behavior seen in Figure~\ref{fig:fesc_all_1} is determined by the ordering of these timescales 
 within the shaded transport regimes. Neutron-mediated escape is efficient only when a parent CR proton is likely to undergo a hadronic interaction before escaping by diffusion as a charged particle, and when the resulting neutron can traverse the system before decaying. In the galactic models, especially the starburst case, the onset of pp interactions arises at relatively low energies (from $\sim$10$^9$ eV). However, the corresponding neutrons decay too quickly to allow passage out of the system (unless the neutron was formed in the vicinity to an escape surface). This means that neutron-inclusive and no-neutron escape fractions remain similar over most of the energy range. 
 By contrast, in clusters and filaments the charged-particle escape time remains long, while 
 at ultra-high energies (above $\sim$10$^{20}$ eV), the neutron decay time eventually exceeds the relevant crossing time. 
 In this regime, temporary charged--neutral switching re-opens an escape channel 
 that would otherwise be strongly suppressed by magnetic confinement and interaction losses, producing 
 the excess escape evident in Figure~\ref{fig:fesc_all_1}. 

\begin{figure}[H]
{\captionof*{figure}{where the gyro-radius exceeds the distance to the nearest confinement boundary. Thus, CRs in blue-shaded regions are under magnetic confinement, with their propagation mediated by diffusive leaking (or neutron-channel escape). CRs in the orange-shaded regions are not magnetically confined and can escape ballistically from the structure (over timescales indicated by the dotted gray horizontal lines), and particles in the unshaded bands (where present) are partially confined depending on the location of the nearest escape surface. Neutron-mediated escape becomes important only where hadronic interactions occur before charged particles escape, and where the neutron decay time exceeds the relevant ballistic crossing time. This accounts for the neutron channel being weak in the galactic models over most of the plotted range, but becomes important at ultra-high energies in the larger-scale cluster and filament environments. Timescales are all calculated for present \mbox{($z=0$) conditions.}}}
\end{figure}

\subsubsection{CR Cascades and Secondaries}
\label{sec:cascades}

\begin{figure}[H]
\centering 
{\includegraphics[width=0.9\textwidth]{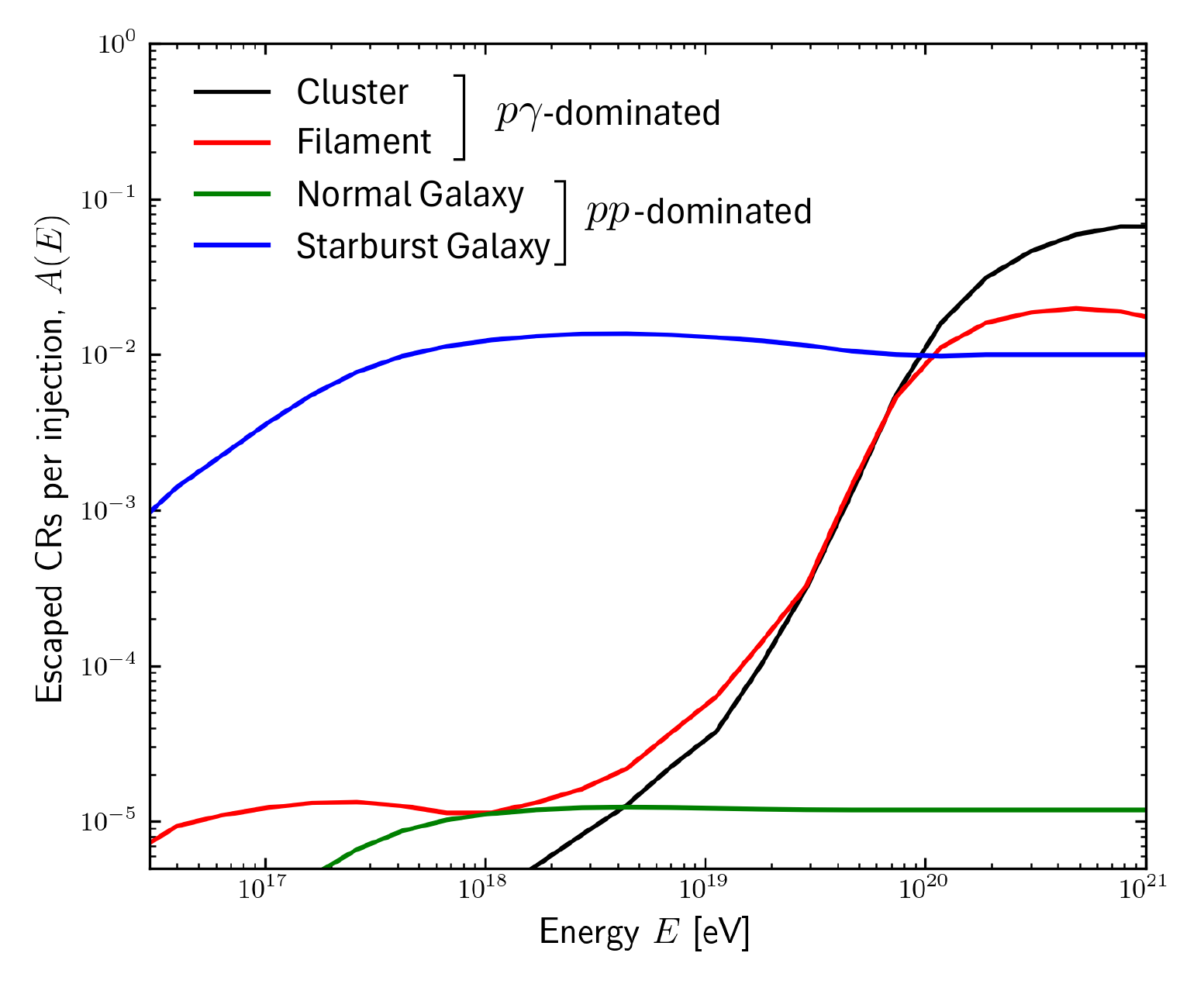}} 
\caption{Escape factor per injected CR proton, $A(E)=N^{n}_{\rm esc}(E)/N_{\rm inj}^{\rm pri}(E)$, showing the mean number of CR nucleons that escape via the neutron-mediated channel per injected primary CR proton at a given energy $E$ for the illustrative model environments: galaxy cluster (black), cosmological filament (red), normal Milky Way-like galaxy (green), and starburst galaxy (blue). Values of $A\ll 1$ indicate that neutron production has little bearing on particle escape (few neutron-channel escapes per injected CR).}
\label{fig:escape_A}
\end{figure} 

The neutron escape channel 
  is not a simple transport process. 
As it hinges on the operation 
  of a hadronic interaction, 
  multiplicities of secondary nucleons 
  (and anti-nucleons) can arise. 
This is quantified by the escape factor per injection, $A(E)$, shown in Figure~\ref{fig:escape_A}, which 
measures the mean number of neutron-channel escapes per injected CR primary in each structure\endnote{Note that the energy range displayed is restricted, such that all systems have reached a quasi-steady state at the simulation ages (cf. diffusion timescales compared to snapshot times shown in Figure~\ref{fig:timescale_fig_1}).}. 
{This gives} an intuitive quantification of the confining capability of the host structure {and the importance of the neutron-escape channel}. 
For $A\ll1$, the structure behaves as an effective CR container. Only a small fraction of injected primary CRs 
contribute to neutron-mediated leakage. For \emph{A(E)}$\sim$1, the structure allows substantial escape via the neutron channel at that energy.  
$A(E)$ is boosted when both the magnetic confinement capability of a structure 
and the hadronic interaction efficiency are high. This is because interactions occur before the CR particles escape, and repeatedly convert baryons into ballistic segments.

\begin{figure}[H]
{\captionof*{figure}{Values of $A(E)$ that are closer to 1 indicate that the neutron escape channel operates more efficiently. The onset energy and growth of $A(E)$ differs by environment, occurring at lower energies in dense galaxies {(compact, pp-dominated systems)} and only at ultra-high energies in clusters and filaments {(extended, p$\gamma$-dominated systems)}.}}
\end{figure}

\subsubsection{Escape Directions}
\label{sec:escape_directions}

\begin{figure}[H]
\begin{adjustwidth}{-\extralength}{0cm}
\centering
{\includegraphics[width=9.0cm]{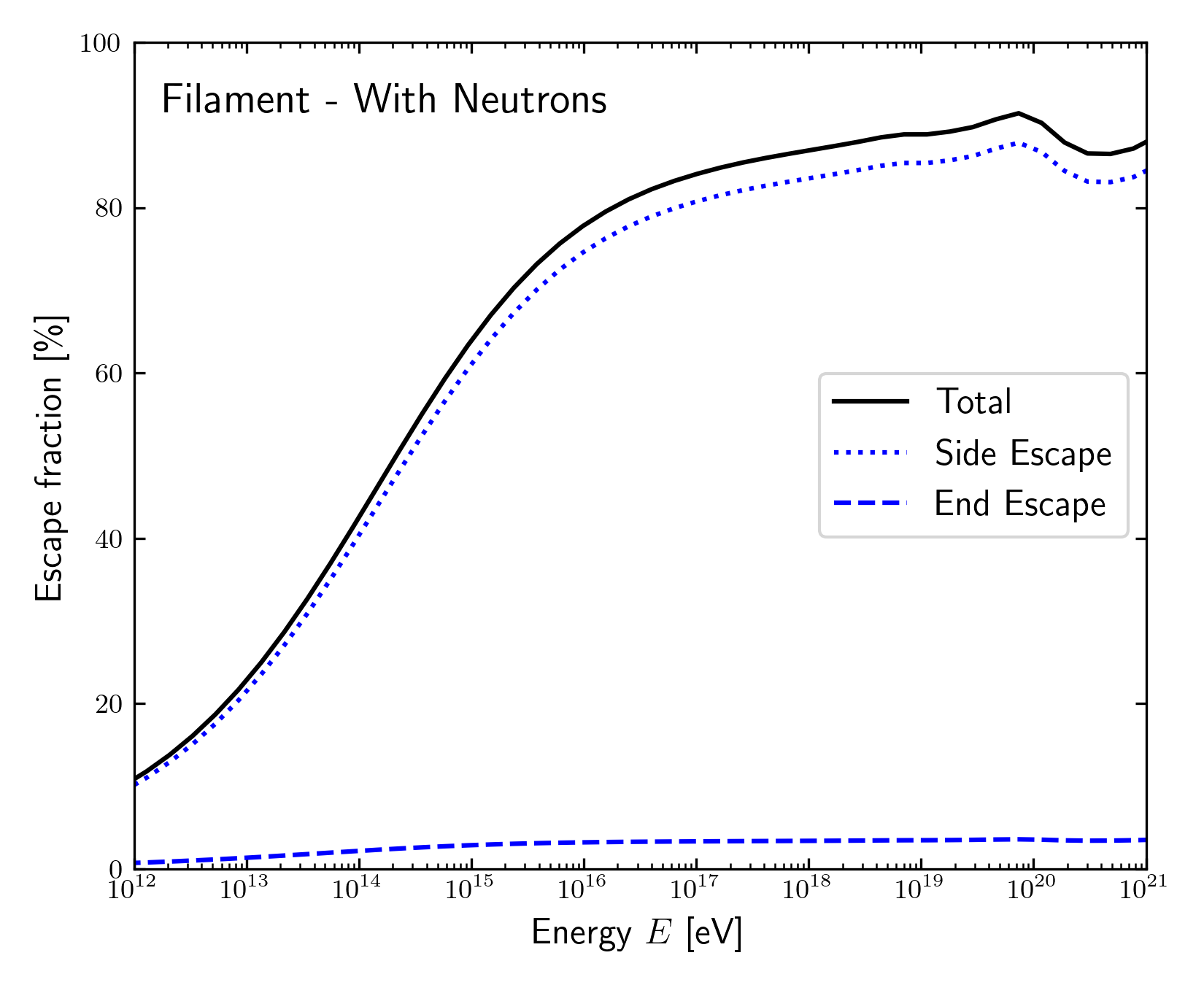}}\\
{\includegraphics[width=9.0cm]{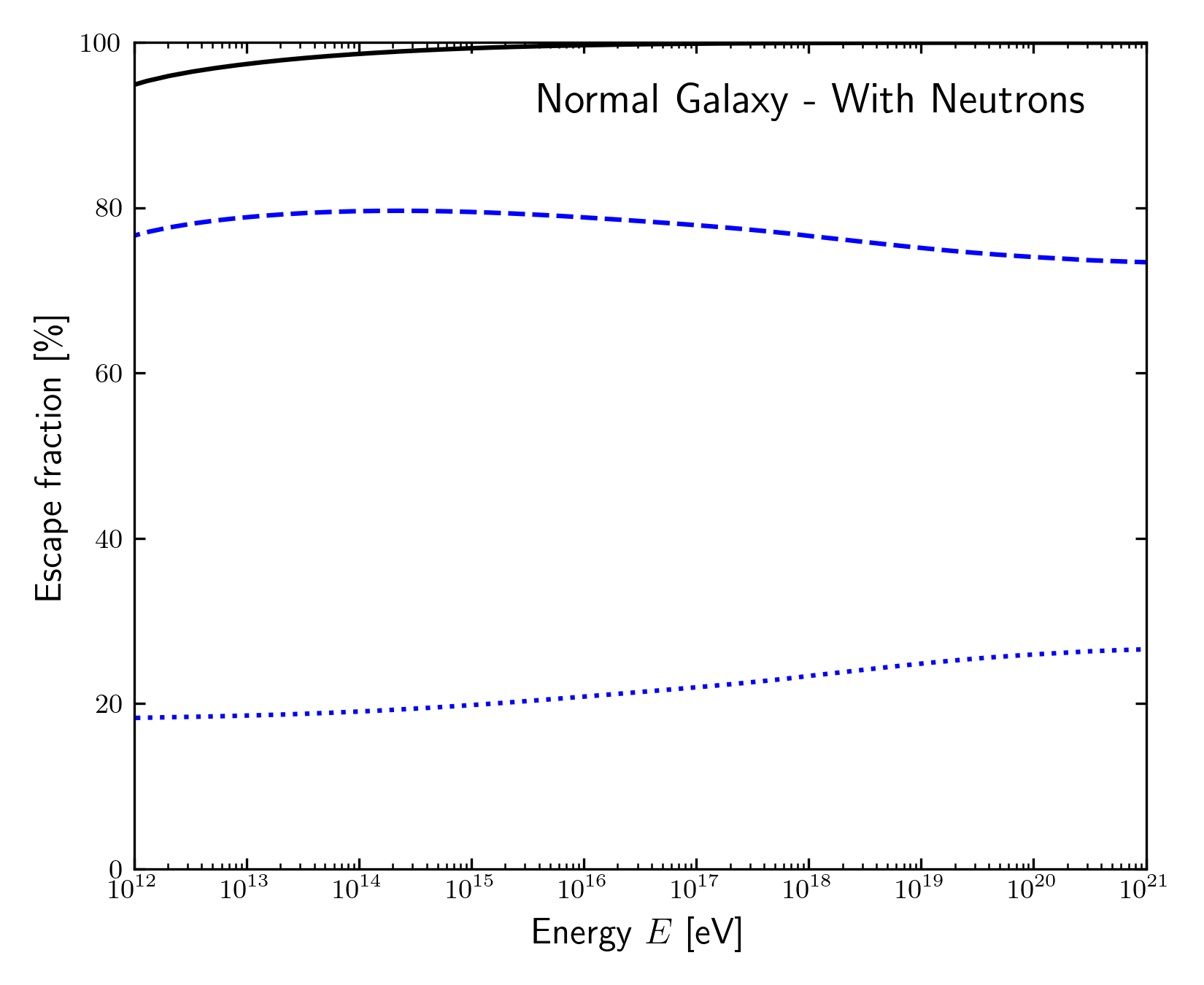}}
{\includegraphics[width=9.0cm]{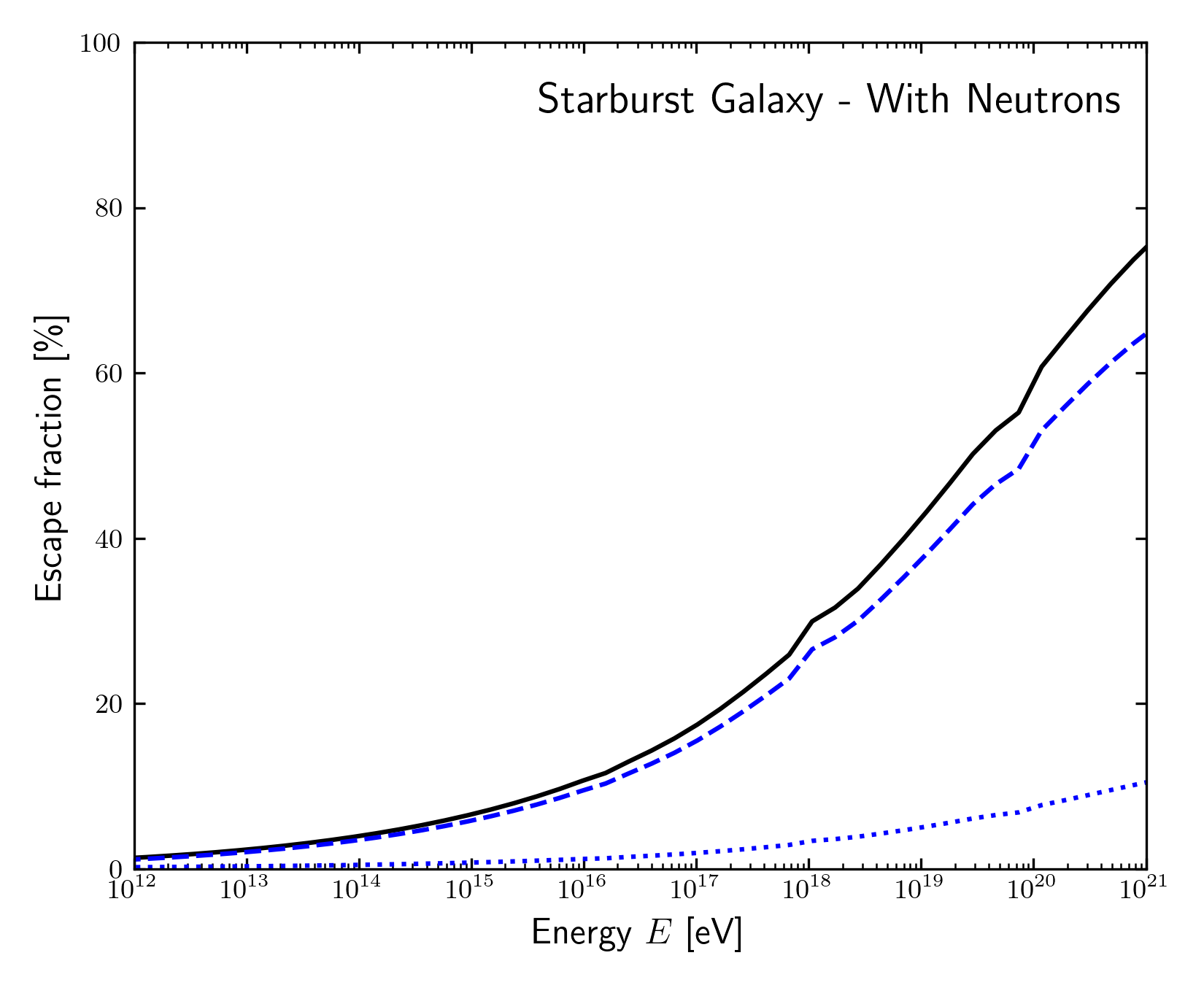}}
\end{adjustwidth}
\caption{Directional decomposition of CR escape from the Milky Way-like galaxy model, where charged--neutral switching is included. The total escape fraction $f_{\rm esc}(E)$ (black) is decomposed into contributions through the cylindrical side boundary at the edge of the galaxy (``side escape'', dotted), and through the end-caps above and below the galaxy's plane (``end escape'', dashed). In all cases, the effect of charged--neutral switching has no major bearing on the escape direction. {This shows how the geometry of the system imprints an anisotropy on the emergent CR flux.}}
\label{fig:galaxy_geometry}

\end{figure}

Figure~\ref{fig:galaxy_geometry} shows the directional decomposition of the emergent CR flux in the cylindrical models. 
This is not strongly modified by the neutron-mediated escape channel. 
As the CR neutrons produced in hadronic processes inherit the direction of their parent CR protons, the inclusion of the 
neutron escape channel would not be expected to strongly re-route the emergent CR flux from a structure. 
However, the escape in general is anisotropic and energy-dependent, 
illustrating the competition between geometry, 
the distance to the nearest escape surface and the efficiency of transport towards each boundary (including any diffusion anisotropy). This indicates 
that magnetized structures can act as directional sieves. They 
regulate the directions CRs of different energies are preferentially released, and how these CRs are seeded into higher-order structures. 

In the flattened galactic geometries, escape proceeds predominantly through the end-caps over most of the energy range we consider. This is because the vertical scale height is much smaller than the radial extent of the disk. The top and bottom surfaces therefore provide the 
shortest path to escape for much of the confined CR population. As the CR energy increases, the 
side contribution gradually increases. This is because higher-energy particles diffuse more 
effectively throughout the full galaxy volume and are less strongly biased towards the 
nearest escape surface.

The elongated filament geometry shows the opposite directional preference. Because the transverse distance to the 
cylindrical side boundary is much smaller than the full filament length, most escaping CR nucleons leave through 
the side surfaces (despite our model invoking strongly anisotropic diffusion along the filament length). The end-cap 
contribution increases modestly at the highest energies, consistent with the faster diffusive transport of the parent CR proton population along the filament spine, but not enough to overtake the side channel.  

This directional energy-dependent sieve effect 
could be non-negligible, especially 
for CR transfer 
between 
hierarchical structures confining CRs. Flattened galaxies preferentially vent CRs out of their planes. 
In contrast, filaments tend to leak CRs laterally into their surroundings. At high energies, this can 
overcome their capability to act as 
 ``highways'' which channel CRs between the clusters they connect to. This indicates that cosmological filaments may   
act effectively as ``pipes'' of lower energy CRs, but those pipes become prone to leaking CRs of high energies. 
The energy dependence of this directional partitioning implies that the angular distribution of 
CR injection into adjacent structures may encode the geometry and transport regime of the host environment, rather 
than simply tracing CR source locations.

\subsubsection{Redshift Dependency}
\label{sec:redshift_dependency}

\begin{figure}[H]
\begin{adjustwidth}{-\extralength}{0cm}
\centering
{\includegraphics[width=9.0cm]{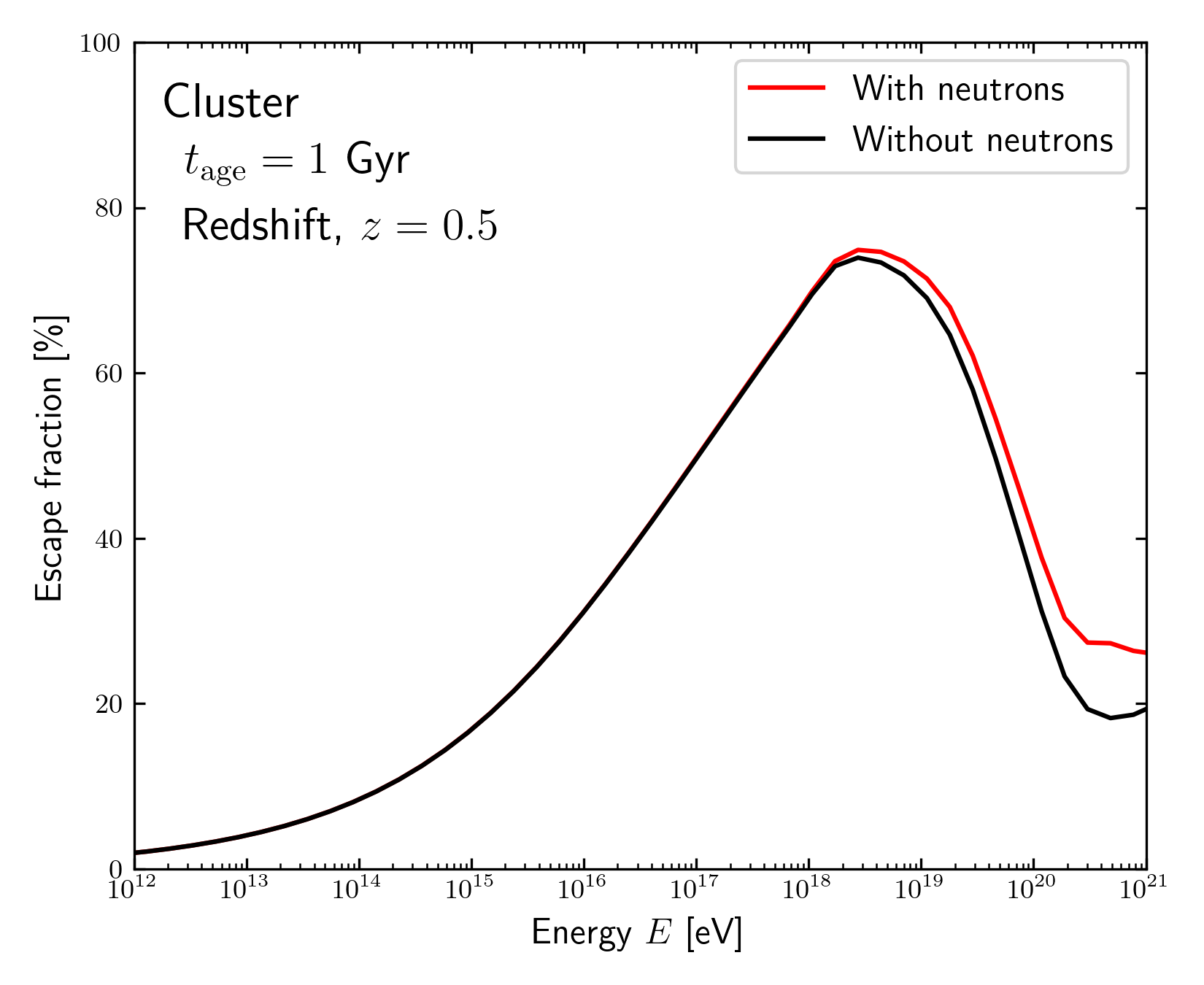}}
{\includegraphics[width=9.0cm]{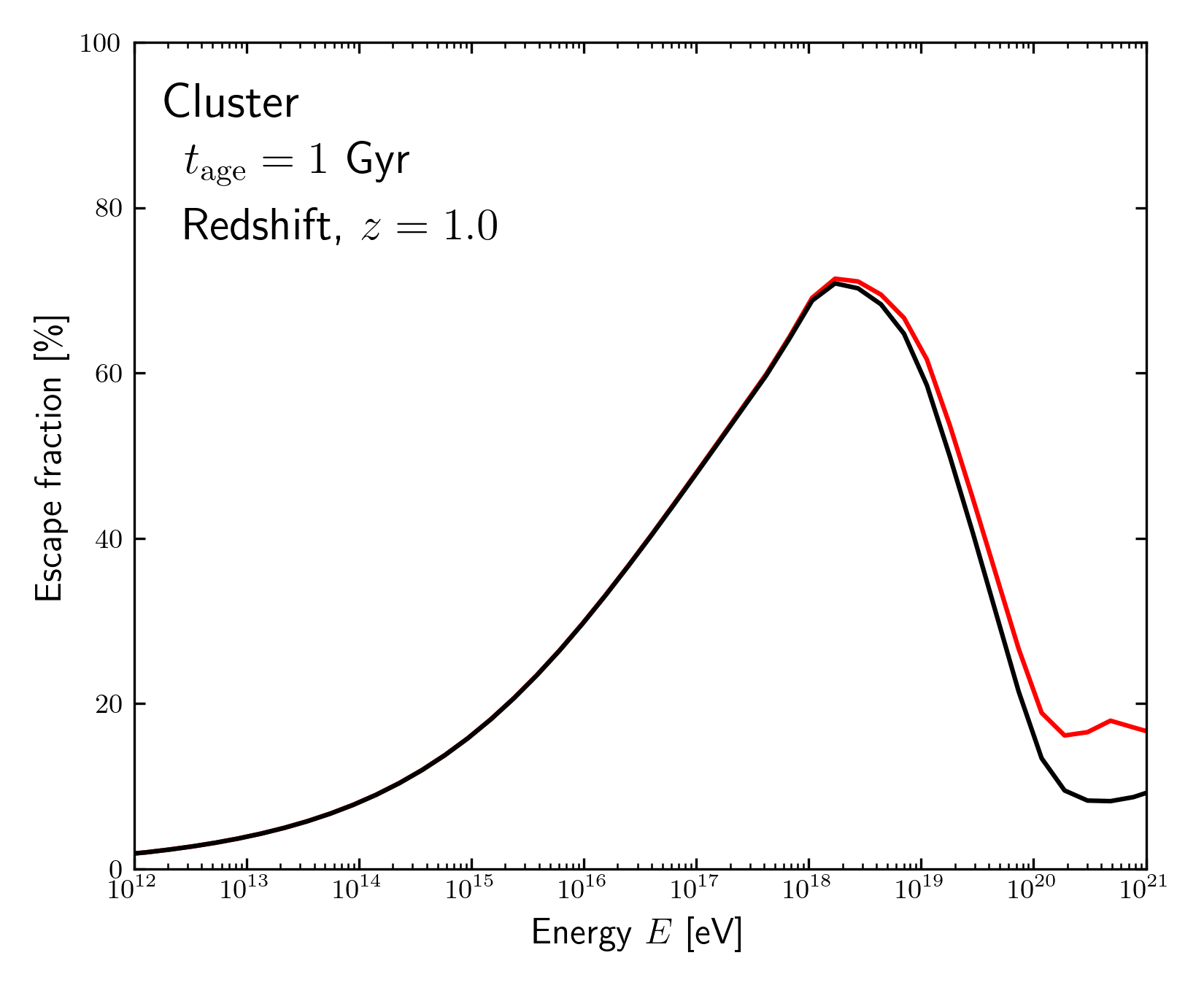}}
\end{adjustwidth}
\caption{Energy-dependent escape fraction, $f_{\rm esc}(E)$, of injected CR protons from the galaxy cluster model 
at higher redshift, shown for $z=0.5$ and $z=1$. Black curves show the standard treatment where hadronic interactions act 
only as sinks for the charged CR population (``without neutrons''), while red curves include charged--neutral switching and neutron-mediated escape (``with neutrons''). Compared with the current epoch results ($z=0$) shown in Figure~\ref{fig:fesc_all_1}, the high-energy turnover shifts to lower energies at higher redshift because p$\gamma$ absorption and cooling losses are more effective. The neutron channel partially restores the escaping ultra-high-energy tail but does not eliminate the overall redshift-driven suppression in CR escape.}
\label{fig:fesc_clus_z}

\end{figure}

\begin{figure}[H]
\begin{adjustwidth}{-\extralength}{0cm}
\centering
{\includegraphics[width=9.0cm]{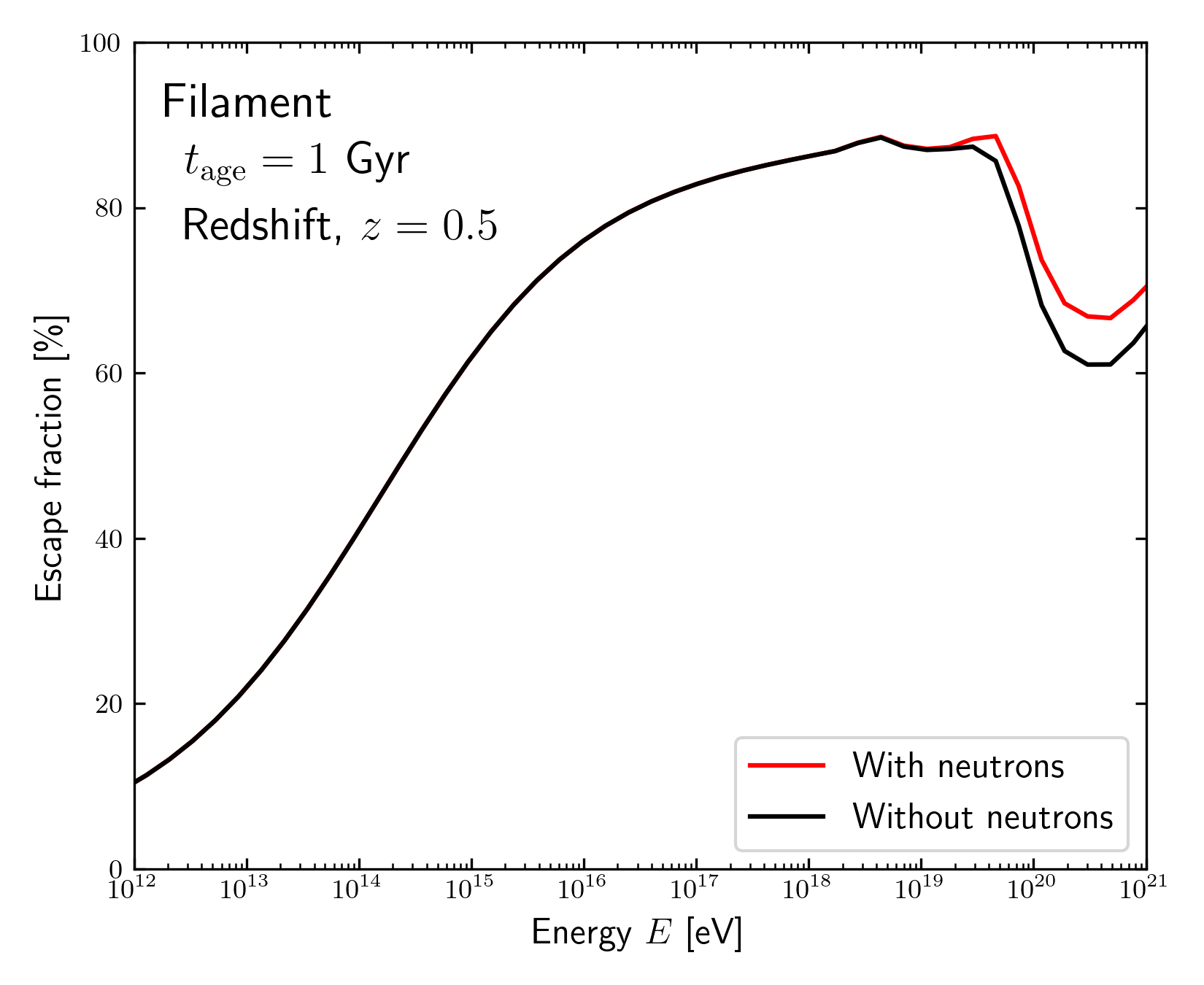}}
{\includegraphics[width=9.0cm]{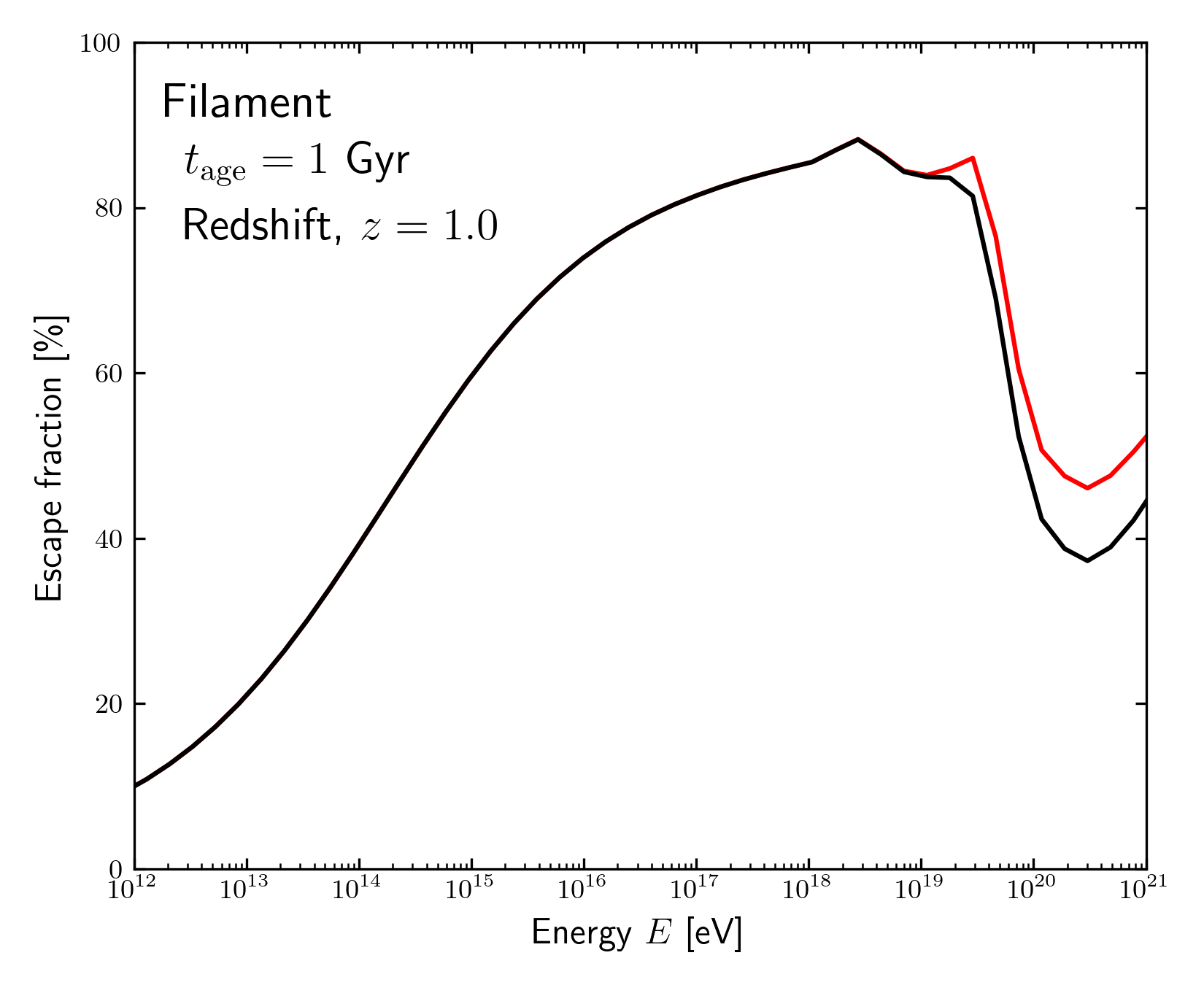}}
\end{adjustwidth}
\caption{Same as Figure~\ref{fig:fesc_clus_z}, but for the filament model. As filaments formed earlier in the Universe, these effects could be extended to earlier epochs, where the neutron-mediated escape channel would provide an increasingly important bypass to the p$\gamma$ attenuation losses seen at ultra-high-energies.}
\label{fig:fesc_fil_z}

\end{figure}

\begin{figure}[H]
\begin{adjustwidth}{-\extralength}{0cm}
\centering
{\includegraphics[width=9.0cm]{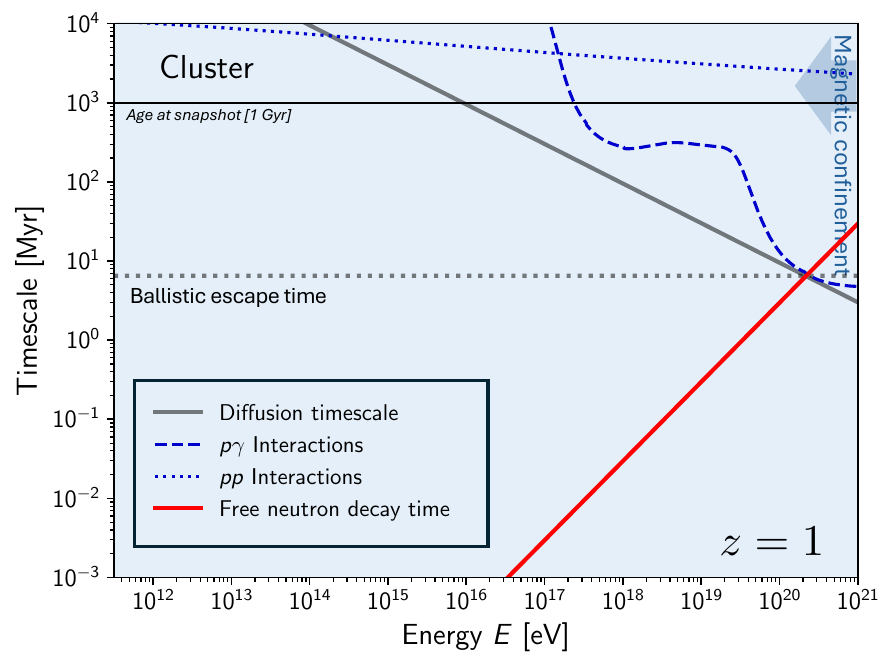}}
{\includegraphics[width=9.0cm]{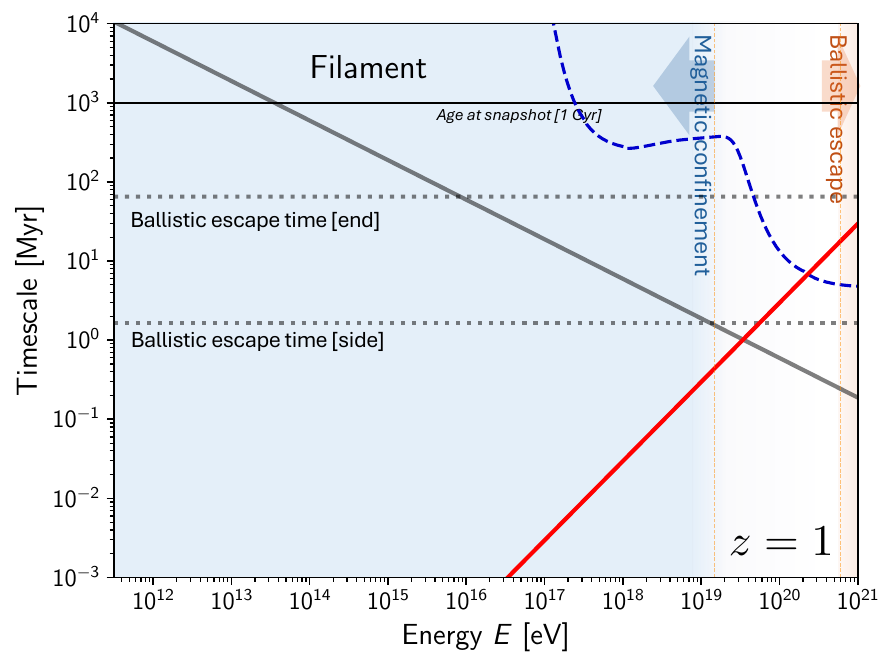}}
\end{adjustwidth}
\caption{Characteristic timescales, same as Figure~\ref{fig:timescale_fig_1}, but governing CR escape in the cluster and filament models at $z=1$.}  
\label{fig:timescale_fig_z1}

\end{figure}

Figures~\ref{fig:fesc_clus_z} and~\ref{fig:fesc_fil_z} show that the role of neutron-mediated escape 
becomes increasingly sensitive to redshift in the large-scale cluster and filament environments. 
As redshift increases, the denser and hotter CMB provides a more effective target field for p$\gamma$ interactions, shortening the corresponding interaction timescale and shifting the onset of strong high-energy attenuation to lower CR energies. This can be understood with the timescale comparison shown in Figure~\ref{fig:timescale_fig_z1}, which illustrates that, at the highest energies, 
the p$\gamma$ interaction timescale becomes more comparable to the crossing times of the confining structures, while the particles remain in the confined (blue-shaded) or partially confined (white) regimes. 
In a purely charged-particle treatment, the proton escape fraction therefore turns over earlier and declines more steeply at the highest energies.  
When charged--neutral switching is included, part of the baryon number removed from the confined proton population is transferred into neutrons, which can then propagate ballistically and potentially leave the system before undergoing $\beta$-decay. 
This channel raises the effective escape fraction above $10^{20}$ eV by $\sim$50$\%$ in galaxy clusters at $z=0.5$, and by around a factor of two at $z=1$. A similar enhancement is also found for filament environments, although for our adopted parameter choices the relative increase is somewhat less pronounced. 
Thus, while neutron production does not eliminate the redshift-induced suppression of CR escape at the highest energies, it does partly offset it by sustaining a residual high-energy leakage beyond the turnover. 

These results suggest that CR transfer across large-scale structure may itself evolve with cosmic time. 
Because filaments emerge earlier than galaxy clusters during structure formation~\cite{Wen2012ApJS, Zhu2021ApJ}, neutron-mediated transport may provide an especially relevant channel for the redistribution of UHE CR baryons through the cosmic web at earlier epochs, before mature clusters become the dominant magnetized reservoirs.  
As the Universe evolves and increasingly massive cluster environments assemble,  
charged-particle confinement and interaction processing become more severe. 
However, the neutron channel can still provide a partial route for the highest-energy baryons to escape.  
We note that, in our model, these changes are brought out by the redshift dependency of the CMB only. The physical properties of clusters and filaments may also vary with redshift, and their evolution may further modify their capability to entrap CRs. This could also affect the relative importance of the neutron escape channel compared to our results. These detailed dependencies are left for dedicated investigation in future work.  

\section{Discussion}
\label{sec:sec4}

\begin{figure}[H]
\begin{adjustwidth}{-\extralength}{0cm}
\centering 
{\includegraphics[width=1.3\textwidth]{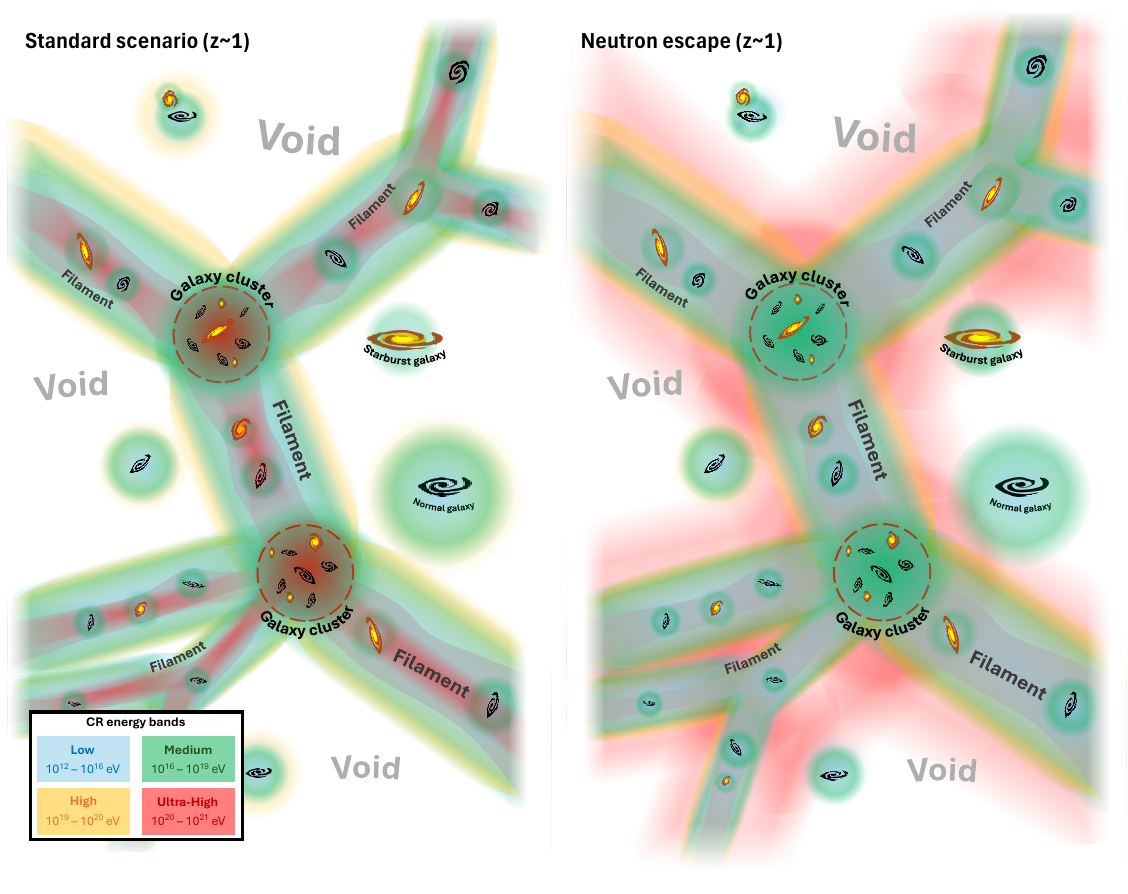}} 
\end{adjustwidth}
\caption{Schematic of neutron-enabled CR leakage across the cosmic web expected for a high-redshift scenario, showing how charged--neutral switching in hadronic interactions can alter the large-scale distribution of escaping CR nucleons across a simplified cosmic-web environment (clusters, filaments, galaxies, and surrounding voids). \textbf{Left}: The ``standard'' propagation scenario, where hadronic interactions act only as sinks for charged CR protons, leading to a more localized CR presence within the magnetized host structures, particularly at high energies. At low and medium energies, CRs can still generally escape from their confining structures over time, but the leakage rate is slow and excesses are expected in the vicinity of these sources. They likely remain magnetically tied to web structures with limited penetration into deep voids. At high and ultra-high energies, CRs can be partially or fully confined by their host structures.  \textbf{Right}: The ``neutron jump'' scenario, allowing neutron production to provide a ballistic bypass and survival channel as explored in this work. Neutrons can cross escape surfaces before decaying, re-seeding charged CRs beyond their host structure and increasing the supply of CRs to lower-density regions (including voids) and across structural interfaces. Colors indicate representative CR energy bands as shown in the legend (\textbf{Low}, blue:~$10^{12}$--$10^{16}$ eV; \textbf{Medium}, green: $10^{16}$--$10^{19}$ eV; \textbf{High}, yellow: $10^{19}$--$10^{20}$ eV; \textbf{Ultra-high}, red: $10^{20}$--$10^{21}$ eV).}  
\label{fig:particle_transfers_schematic}

\end{figure}

Taken together, 
  our results in Section~\ref{sec:results} imply a qualitative 
change in the expected CR exchange between clusters, filaments, and voids when accounting for the production of CR neutrons 
and their ballistic jumps. 
Without explicitly accounting for the 
effect of these stochastic CR neutron jumps, the highest-energy CRs are strongly attenuated in environments where interaction times are shorter than escape times, curtailing the supply of ultra-high-energy CR nucleons into larger-scale reservoirs. However, 
with neutron-mediated escape, the severity of this attenuation and confinement is weakened at the highest energies, allowing more ultra-high-energy CR nucleons to contribute directly to the the CR content outside their host structure. 
This provides a natural mechanism for seeding filaments and voids with high-energy CR nucleons, sourced from connected lower-order astrophysical structures. The resulting picture is illustrated schematically in Figure~\ref{fig:particle_transfers_schematic} for a high-redshift scenario (where the impact of the neutron escape channel would be more pronounced than in the present epoch). 

\subsection{Further Remarks and Astrophysical Implications}
\label{sec:further_implications}

Beyond the transfer of CR baryons across the hierarchy of cosmological and astrophysical ecosystems, our work highlights that hadronic interactions need not act purely as an absorbing sink for a magnetically confined CR population. When charged--neutral switching is considered, hadronic interactions provide a ``ballistic bypass'' channel that partially decouples the fate of injected ultra-high-energy CRs from magnetic confinement 
and severe photo-attenuation. 
In environments where a standard absorption-only treatment predicts strong suppression due to hadronic interaction timescales 
being shorter than charged particle escape timescales, the neutron channel can restore 
an escaping component by converting confined   
protons into free-streaming nucleons that cross escape surfaces before decaying. This picture has several implications for  (i) CR transport through cosmological filaments, and (ii) the viability of cluster environments as contributors to the ultra-high-energy CR flux observed at Earth.  

\subsubsection{CRs in Cosmological Filaments: CR Highways or Leaky Pipes?}
\label{sec:highways_or_pipes}

Cosmological filaments are large-scale magnetized structures that sit at the top of the cosmic hierarchy. They host ordered large-scale magnetic fields oriented along their extension which have been previously been discussed as potential ``highways'' that guide CRs efficiently along the cosmic web~\cite{Kim2019SciA, Wu2024Univ}. This picture is aided by anisotropic diffusion operating more favorably along the direction parallel to the background large-scale magnetic field vector, aligned with the filament axis~(e.g., \cite{Banfi2021MNRAS,Carretti2022MNRAS,Balboni2023A&A, Vernstrom2023SciA,Carretti2025Univ}). However, our directional escape results suggest a more nuanced picture. For an elongated filament geometry, CR leakage is typically dominated by losses through filament walls, with only a portion of the CR population being channeled along the filament to its ends---and only at energies below $\sim$PeV energies, where propagation is diffusion dominated (see Figure~\ref{fig:galaxy_geometry}). This suggests that these structures may be more correctly considered as leaking CR ``pipes'' rather than true CR ``highways''. 

Our results imply that filaments still operate as organizing structures for CRs, even while being efficient in venting a large fraction of their CR content into surrounding lower-density regions (voids). In the neutron-enabled scenario at ultra-high energies, this venting can include ultra-high-energy CR nucleons that would 
be considered partially attenuated in a purely charged treatment, particularly during earlier cosmological epochs. This alters the picture for CR penetration into the filament outskirts and adjacent void environments. 

\subsubsection{Escape of CRs from Galaxy Clusters}
\label{sec:cr_escape_clusters}

Galaxy clusters have long been considered CR reservoirs, up to ultra-high energies. 
They host ample candidate source environments (supernovae in galaxies, accretion shocks, AGN and associated jets or radio lobes~\cite{Murase2008ApJ, Kotera2009ApJ, Harari2016JCAP, Fang2018NatPh}; see also Refs.~\cite{AlvesBatista2019FrASS, Globus2025ARA&A} for recent reviews) and their moderately strong ($\upmu$G-level), turbulent magnetic fields imply long CR confinement times exceeding Gyrs~(e.g., \cite{Harari2016JCAP}), even up to extremely high energies~\cite{Fang2016ApJ}. 
The CRs trapped inside clusters are often 
regarded to have difficulty reaching us on Earth, as they can be efficiently 
attenuated by hadronic p$\gamma$ interactions in the intra-cluster environment, meaning that clusters may be efficient ultra-high-energy CR \mbox{calorimeters~(e.g., \cite{Kotera2009ApJ,Harari2016JCAP,Condorelli2023ApJ}).} In this picture, CRs are rapidly reprocessed into secondary particles and emissions~(e.g., \cite{Murase2008ApJ,Kotera2009ApJ,Fang2018NatPh}). However, recently, some authors have claimed excesses of ultra-high-energy CRs from the directions of nearby clusters/super-clusters~\cite{TA2021arXiv, Kim20253W}, and have found correlations with the matter distribution of the local Universe~\cite{Ding2021ApJ}. 

Our results 
indicate that it is possible for galaxy clusters 
to host sources 
  whose ultra-high-energy CRs could be observable at Earth. 
In the neutron-jump propagation scenario, 
CR neutrons produced stochastically in hadronic interactions propagate ballistically and can escape from galaxy clusters at the highest energies. This provides a ballistic bypass of magnetic confinement which can sustain a substantial escaping CR nucleon flux even in the ultra-high-energy regimes where charged CR primaries are expected to be attenuated by p$\gamma$ interactions in the cluster environment. Our findings relax the requirement that CRs must avoid strong interaction environments in order to escape from sources embedded in galaxy clusters. Such systems therefore remain a potential source class for the observed flux of ultra-high-energy CRs. 

In the context of the inter-connected cosmic-web, this escape mechanism also supports a more efficient transfer of CR nucleons from clusters into their connecting filaments. In particular, our results suggest that clusters can supply larger-scale reservoirs with CRs, and that they 
may ``light-up'' cosmic-web connections as 
proposed by Ref.~\cite{Kim2019SciA}. While Ref.~\cite{Wu2024Univ} demonstrated that this process may be hampered by the presence of magnetic barriers operating to frustrate charged particle exchange at cluster--filament interfaces, CR neutron production could allow such barriers to be bypassed to enable CR passage into nearby filaments, where the filament magnetic fields could re-capture the charged particles. Our results suggest that this bypass mechanism may be more effective in high-redshift settings at an earlier stage of cosmological structure formation. 

\subsubsection{Comment on CR Confinement and Calorimetry in Our Results}
\label{sec:cr_calorimetry}

We have found that CR escape from the adopted fiducial model for a ``normal'' galaxy 
is very efficient, while 
strong confinement arises in our ``starburst'' model (see Figure~\ref{fig:fesc_all_1}).  
CR confinement and calorimetry in galaxies are often discussed in terms of the attenuation of the hadronic CRs and 
their deposition of energy and momentum 
  (e.g., \cite{Owen2023Galax} for a review). 
At GeV--TeV energies, 
  the dominant process for hadronic CRs
  is inelastic pp collisions in the dense ISM. 
Calorimetry is often framed in terms of the ability of 
  these CRs to influence the host galaxy ecosystem 
  through heating, ionization, and momentum injection, 
with signatures such as $\gamma$-rays and neutrinos  
  being diagnostics of energy \mbox{deposition~e.g.,~\cite{YoastHull2015MNRAS,Werhahn2021MNRAS}).} 
Strong starbursts are therefore expected 
  to approach hadronic calorimetry, 
  whereas Milky Way-like galaxies  
  are comparatively ``leaky''. 

In its simplest form, 
  hadronic calorimetry can be reduced 
  to a timescale criterion. 
A system is calorimetric when the characteristic loss time of its CRs, $\tau_{pp}$, is 
shorter than the effective CR escape time, $\tau_{\rm esc}$. 
The loss time is set by the density of the ISM, 
   the main target for hadronic collisions, i.e., $ \tau_{\rm pp} = 1/({n \;\! \sigma_{\rm pp} \;\! c})$, where 
$ \sigma_{\rm pp}\approx$ 30  mb above the pp interaction threshold (see Section~\ref{sec:hadronic_interctions}), with limited energy dependence~\citep{Kafexhiu2014PhRvD}. 
If escape is dominated by diffusive leaking, $\tau_{\rm esc}$ can be estimated as the diffusive crossing time across the characteristic vertical scale height of the galaxy, $\tau_{\rm esc} \simeq {{H_{\rm gal}}^{2}}/{2\,|\mathbf{K}|}$, where the diffusion parameter $\mathbf{K}$ is a set by the CR energy $E$ and effective mean magnetic field strength, $\langle|\mathbf{B}|\rangle$. 

Adopting galaxy parameter values from Tables~\ref{tab1:physical_params} and ~\ref{tab:geometrical_params} for our 
characteristic ``starburst'' and ``normal'' galaxy models, 
and assuming that the physical conditions within them are spatially uniform, the loss 
timescale for the ``normal'' galaxy for 1 TeV CR protons is 
$\tau_{\rm pp} \approx 30 \;\! {\rm Myr} $, 
and the escape timescale is $\tau_{\rm esc} \approx 1 \;\! {\rm Myr}$. For CRs of the same energy in the characteristic ``starburst'' galaxy model, these timescales are instead $\tau_{\rm pp} \approx 0.1 \;\! {\rm Myr}$ and $\tau_{\rm esc} \approx 10 \;\! {\rm Myr}$.  This shows that, for the specific parameter choices adopted here, TeV hadrons are expected to escape efficiently 
in our normal galaxy model, 
while they will be strongly attenuated in starburst galaxies (see also Figure~\ref{fig:timescale_fig_1}). This is consistent with 
 the established picture that Milky-Way like (``normal'') galaxies are non-calorimetric, while starbursts can be highly calorimetric, with a sensitive energy-dependence of CR confinement developing from the competition between pp absorption and diffusive transport. The charged--neutral switching effect 
considered in our work does not significantly alter this GeV-TeV calorimetry argument. 
At the energies of interest (and below), neutron decay lengths are far smaller than galactic scales. CR neutrons do not provide a direct ``ballistic bypass'' of confinement comparable to the cluster/filament case. 

\subsubsection{CR Injection Timescales and Transient Sources}
  \label{sec:transients}

In our calculations, we considered that CRs are injected by a persistent, non-evolving source population embedded 
within each structure. The escape fractions we obtain can therefore be 
interpreted as long-term energy-dependent transfer efficiencies for CR baryons 
at a given epoch. In reality, 
the CR sources in galaxies, clusters and filaments would be time-dependent. 
In galaxies, the accelerators of the CR population are expected to be associated with 
star-formation activity, or with processes related to AGN activity~(for reviews, \mbox{see~\cite{Owen2023Galax,Ruszkowski2023A&ARv}).} 
Galaxy-scale star-formation episodes typically persist over hundreds of Myr~\cite{DiMatteo2008A&A} (i.e., longer than our simulation time), although shorter-timescale flickering associated with individual star-forming clouds occurs on scales on the order of tens of Myr~\cite{Chevance2020SSRv}. 
By contrast, AGN variability can occur on much shorter timescales, on the order of \mbox{0.1 Myr~\cite{King2015MNRAS, Schawinski2015MNRAS}. } 

Similarly, the CR supply in clusters of galaxies may be dominated by repeated AGN activity from a brightest central galaxy, with individual outbursts or recurrence cycles arising on 5--10 Myr timescales. 
Moreover, the resulting 
CR population may be dynamically redistributed by bubbles and weak shocks over a few tens of Myr. Subsequent cavity evolution is commonly characterized in terms of sound-crossing, buoyancy, and refill timescales, while transport and possible re-acceleration associated with sloshing motions can persist over $\sim$Gyr timescales~\cite{Rafferty2006ApJ, Sanders2016MNRAS, Bourne2023Galax}. 
The origins of CRs in filaments are less certain but likely include in situ acceleration at structure-formation and accretion shocks~\cite{Ryu2003ApJ, Vernstrom2023SciA}, 
as well as injection from lower-order structures through AGN activity and feedback, for example from giant radio jets~\cite{Oei2024Natur, Oei2024A&A}, 
and weaker delivery channels such as lower-power AGN outflows or stellar/galactic winds from filament galaxies~(e.g., \cite{Bertone2006MNRAS, Gheller2015MNRAS, Zhang2018Galax, Vazza2025A&A}). The characteristic timescales of these processes may differ substantially, potentially even experiencing  
variability over the Gyr timescales associated with filament evolution. 

When the variability timescale of CR sources is substantially shorter than the evolutionary timescale of the structure confining the CRs they produce, successive bursts populate the reservoir faster than the stored high-energy CR population is depleted. 
In this situation, the escaping component approaches the quasi-steady, time-averaged behavior captured by our calculations (at higher energies). 
We consider this to be a plausible scenario in large-scale reservoirs, such as galaxy clusters and filaments, where the relevant charged-particle confinement times can be extremely long, while the neutron-mediated channel operates through ballistic crossings over much shorter timescales. The structure then acts as a time-integrated fossil record 
of multiple source episodes, and our reported escape fractions offer a reasonable characteristic description of the cumulative leakage.  

If the duty cycle of sources is low, and bursts of CR injection are separated by 
intervals comparable to or longer than the depletion or evolution time of the high-energy confined 
CR reservoir, then the escaping component should retain a memory of the injection history rather than approaching a time-averaged state. 
In this situation, neutron-mediated escape may be especially important 
because it can produce a comparatively prompt high-energy response during active phases.  
In contrast, the more strongly confined charged component can continue to leak out more gradually after the source has faded. 
A fully time-dependent treatment would therefore rescale the normalization of the escaping 
flux from a structure, and also introduce intermittency, spectral hysteresis and environment-dependent 
delays between source activity, in situ secondary production and the escaping CR baryon flux. 
Extending the framework introduced in this study to episodic injection histories 
is therefore a natural future step, and may help to identify the source 
classes, duty cycles and environments where 
charged--neutral switching leaves the clearest observable imprints on CR transport 
through galaxies, clusters and the cosmic web.  

\subsubsection{Direct and Indirect Observational Prospects}
  \label{sec:direct_obs}

  A direct observational test of the escaping CR neutron component considered in this 
  work is challenging. Even at an energy of 10$^{21}$ eV, free neutrons travel only a distance on the order of a few Mpc before undergoing $\beta$-decay, 
  so the results presented here do not imply that neutron point sources from galaxy clusters or cosmological filaments could be detected at Earth.  
  Instead, they imply a modification to the escape efficiency of CR baryons from such environments  
  and, hence, to the charged CR component that is ultimately supplied to the wider cosmic web ecosystem after neutron decay.  
  This interpretation is consistent with the recent search by the Pierre Auger Observatory for ultra-high-energy 
neutrons from more than 1000 galactic candidate sources, which found no significant excess and placed upper limits 
on the neutron flux, even in the regime where galactic neutron astronomy would, theoretically, be possible~\cite{PAO2026arXiv}. 

{More promising observational consequences lie in anisotropy signatures in the UHE 
CR sky. In particular, correlations between UHE CR arrival directions and nearby
large-scale structure would be sensitive to the propagation physics of CRs, including neutron-channel jumps. 
In the nearby-source regime relevant above $10^{19}$ eV (i.e. within photo-attenuation horizons), where
intergalactic magnetic-field deflections are not consequential, temporary neutral
segments in UHE CR propagation can modify how efficiently CR baryons escape magnetized
clusters and filaments. Relative to a purely charged-particle treatment, this would alter the energy dependence,
characteristic angular scale, and rigidity dependence of correlations between the
highest-energy CR arrival directions and nearby large-scale structure. In our
calculations, this is particularly relevant for clusters (and for filament regions
adjacent to clusters), where neutron-mediated escape modifies the local-Universe escape
fraction at the level of up to $\sim$10 per cent (Figure~\ref{fig:fesc_all_1}). 
Although modest in amplitude, such changes are still capable of producing distinctive
broadening or smearing in angular-correlation studies with nearby matter tracers.}

Anisotropy correlation searches of exactly this kind are already standard in the field, including large-scale
dipole analyses, intermediate-scale excess searches, comparisons with nearby galaxy
catalogs, and harmonic cross-correlations with galaxy surveys~(e.g., \cite{Takami2009JCAP,
Das2008ApJ, DelignyPTEP2017, Urban2021A&A, Tanidis2022JCAP, Urban2024OJA}). 
Our findings in this work therefore identifies charged--neutral switching as a transport ingredient with an 
imprint that may be isolated through the energy- and angular-scale dependence of UHE CR
anisotropy. The 
existence of such an imprint follows qualitatively from the transport effect identified
here. However, a fully quantitative forecast of the exact correlation amplitude, angular scale,
or tracer dependence would require end-to-end mock-sky simulations including
realistic source populations, compositions, and detector exposures. 
This is beyond the scope of the current study, and left to dedicated future work. 

{Composition-dependent anisotropy studies provide an additional diagnostic of 
neutron-mediated CR transport. Magnetic 
smearing of anisotropy signatures depends on CR rigidity.  
Therefore, any transport effect that modifies the escaping nucleon
component should be most cleanly reflected in light or proton-rich subsets of the
highest-energy events, while heavier nuclei would be expected to retain weaker directional
correlations at fixed energy. Such composition-informed anisotropy analyses are now
also beginning to be explored explicitly~(e.g., \cite{Globus2019MNRAS, Allard2022A&A,
Tanidis2024JCAP, PAO2024ApJ})  
and these would have the capability to isolate a neutron-mediated effect when 
sufficient data has accumulated. }
  
{Complementary tests and constraints can be obtained by multi-messenger studies of 
individual sources. } For example, in galaxy clusters, 
the same hadronic interactions that drive neutron production will also generate 
$\gamma$-rays and neutrinos. Neutron production and decay effectively redistributes the baryon energy budget. It  
shifts part of the CR power away from local reprocessing into $\gamma$-rays and neutrinos, and instead channels it into escaping nucleons. 
Existing observational limits already provide useful constraints. For example, a 
stacking analysis of 1094 \textit{Planck} clusters with IceCube found no significant 
neutrino emission~\cite{Abbasi2022ApJ}. Detailed multi-messenger predictions for individual clusters, including CR re-acceleration models for radio halos, as well as for cosmological 
cluster populations, likewise show that the total combined $\gamma$-ray and neutrino output of  
clusters is already constrained by radio, neutrino and $\gamma$-ray data~(e.g., \cite{Nishiwaki2021ApJ,Hussain2021MNRAS,Nishiwaki2023ApJ,Hussain2023NatCo}). 
 Given these constraints, 
 the neutron-mediated escape channel considered in this work would 
 most naturally be inferred by joint modeling rather than through 
 a single direct detection. 
 Relative to a purely calorimetric galactic cluster, enhanced CR 
 escape through temporary neutron jumps would reduce the fraction of CR power reprocessed locally into 
 $\gamma$-rays and neutrinos, while increasing the baryonic CR flux transferred to higher-order 
  environments \mbox{(e.g., connected filaments).} 

\subsection{Approximations and Limitations}
\label{sec:sec_caveats}

Several assumptions 
  and simplifications  have been 
 invoked in our calculations. 
We summarize the key ones in this section 
  and discuss the probable impacts 
  on the findings derived from our calculations. 

\subsubsection{CR Composition}

In our calculations, CR protons 
  can be converted to neutrons in hadronic interactions. 
The actual composition of CRs, particularly at ultra-high energies, 
  is not well constrained. 
Some studies 
  indicate that the contribution of 
  heavy nuclei increases with energy, with proton 
  dominance broadly ruled out~(e.g., \cite{AlvesBatista2019FrASS, Mayotte2025arXiv250710292M, AbdulHalim2025PhRvL, Abbasi2024PhRvD}). 
Interactions of heavy nuclei with radiation fields 
  lead to photodisintegration,  
  with the production of secondary nucleons 
  together with 
  free baryons (protons and neutrons) 
  along their paths. 
Photodisintegration increases the neutron yield 
  for a given CR energy flux.  
It thus boosts the efficiency of neutron-driven escape compared to the results presented in this study. 
A full, composition-dependent treatment of these effects 
  is left to future work, 
  but we note that our main qualitative findings are robust. 
CR composition variation would mainly rescale the effective
propagation lengths and escape efficiencies by a factor of a few, rather than substantially adjusting the overall behavior of CR escape. 

\subsubsection{Neutron Jump and Timescales}

For the confined charged CR particles, 
  we model their transport and population 
  evolution self-consistently. 
To model the neutral particles, 
  we use a ``jump'' approximation. 
After their production, we consider that neutrons 
propagate ballistically over a characteristic
length-scale and then either decay or 
leave the system in a single step. 
This is an appropriate approximation  
  for a CR ensemble evolving on timescales 
  much longer than 
  the time required for a neutron to cross 
  the confining structure, 
  $\tau_{\rm cross} \sim \ell_{\rm str}/c$, where $\ell_{\rm str}$ is
the structure's characteristic size. 
For Mpc-scale filaments or clusters, this 
corresponds to $\tau_{\rm cross} \sim 3~{\rm Myr}$, and for kpc-scale galaxies, $\tau_{\rm cross} \sim ~{\rm kyr}$. Over shorter timescales, a 
full dynamical treatment of the coupled proton--neutron system is 
required to properly track the time evolution of the system. 
Our 
approach is therefore robust over 
timescales greater than $\tau_{\rm cross}$, or in the quasi-steady limit (see also our discussion in Section~\ref{sec:transients}). 

\subsubsection{Magnetic-Field Topology and CR Transport}
\label{sec:magnetic_topology_approx}

The formulation we construct 
for CR transport 
adopts simplified geometries 
and fiducial characteristic diffusion properties to isolate the effect of temporary charged--neutral switching. 
In reality, the magnetic fields of galaxies, clusters and filaments are multi-scale 
and are composed of both ordered and 
turbulent components with spatially varying coherence lengths and topologies~\cite{Carilli2002ARA&A, Beck2015A&ARv, Rajpurohit2022A&A, Vernstrom2023SciA, Carretti2025Univ}. 
The relative importance of these components can have a strong bearing on the effective propagation of CRs, 
including the ratio of parallel to perpendicular diffusion. This can affect the characteristic displacement and escape time of charged CRs~(e.g., \cite{Yan2008ApJ, Xu2013ApJ, Lazarian2023FrASS}). 
As a result, the precise onset energy, normalization and directional decomposition of the 
escape fractions we derive in this work should not be regarded as unique predictions suitable 
for application to all systems of each \mbox{representative type.} 

The explicit consideration 
of these effects would be necessary 
to construct robust descriptions of anisotropic CR transport in specific applications, 
and would be required to obtain exact values for spectral features and escape fractions. 
A broader exploration of the effects of specific magnetic field configurations, coherence 
scales, and diffusion anisotropies is left to future work. However, 
despite our simplified approach, we consider that 
the main qualitative conclusions of this work are robust---i.e., that 
any field-dependent propagation can be bypassed if hadronic interactions 
temporarily convert magnetically continued CRs into neutral particles to allow a 
ballistic transport phase that is not subject to the \mbox{same confinement. } 

\subsubsection{Hadronic Interactions in the Neutron Phase}
\label{sec:neutron_interactions}

A more consequential limitation of our neutron-jump framework is that it does
not include interactions experienced by neutrons during the ballistic ``jump'' phase.
Cooling processes such as adiabatic losses due to Hubble expansion are not
expected to be important over the scales and energies considered in our calculations, but
hadronic interactions of CR neutrons with radiation and matter fields can be non-negligible (np and n$\gamma$ processes; see Equation~(\ref{eq:ng_int})). 
The total n$\gamma$ and np cross sections
are of similar magnitude and energy dependence to p$\gamma$ and pp cross-sections 
up to order-unity iso-spin factors at the energies of interest in this work~(i.e., in the $\Delta$-resonance regime and above; see \cite{Ireland2020PrPNP, Krusche2003PrPNP, PDG2024PR}). Therefore, the
interaction rate experienced by CR neutrons in the radiation fields of our
structures is similar to that of CR protons. 

In our calculations, the neutron channel has the greatest effect at ultra-high 
energies, where (i) the proton interaction length is short and interaction
rates are high, and (ii) the proton gyro-radius is large enough that the
charged particles are only weakly confined and propagate in a quasi-ballistic
regime. In this regime, it is not strictly correct to treat the neutron
propagation as a single, non-interacting jump. Neutrons would 
undergo n$\gamma$
interactions during their flight, similar to their parent protons in the 
confined phase. These interactions both destroy neutrons (producing showers of
secondary CRs) and regenerate new neutrons. This extends the effective distance over which a 
neutral nucleon component is maintained in the CR beam.

The net effect of this neutron cycling can be considered with a simple probabilistic 
estimation. Suppose, after a hadronic interaction, we 
approximate that the leading nucleon is again a neutron with probability $f_n$, and a 
proton with probability $(1-f_n)$. If a proton is produced, the neutron channel is terminated because the proton is confined or treated in the main calculation. 
The number of successive neutron-producing 
interactions, $k$, before the first conversion to a proton then 
follows a distribution with weights
\begin{align}
W_k(f_n) = (1 - f_n) {f_n}^{k-1} 
\end{align}
which are normalized such that $\sum_{\rm k=1}W_k(f_n) = 1$.
The mean number of neutrons produced in this chain per initial neutron is 
\begin{align}
\bar{N}_n &= \sum_{k=1} k W_k(f_n) \nonumber \\
&= (1-f_n) \sum_{k=1} k {f_n}^{k-1} \nonumber \\
&= (1-f_n) \frac{{\rm d}}{{\rm d}f_n} \sum_{k=1} {f_n}^{k} \nonumber \\
&= (1-f_n) \frac{{\rm d}}{{\rm d}f_n}
\left( \frac{f_n}{1-f_n} \right) \nonumber \\
&= (1 - f_n)^{-1}.
\end{align}

This means that n$\gamma$ cycling only changes the number of neutrons by a factor of $\bar N_n = (1 - f_n)^{-1}$, which is of order unity for reasonable values of $f_n$. If we also consider that the final proton terminates the chain, the mean number of total nucleon segments 
in a chain is $\bar{N}_{\rm tot} = \bar{N}_n + 1$. The corresponding neutron fraction of 
the beam path is then 
\begin{align}
    \bar{f}_n &= \frac{\bar{N}_n}{\bar{N}_{\rm tot}} \nonumber \\
    &=\frac{(1-f_n)^{-1}}{(1-f_n)^{-1} + 1} \nonumber \\
    &=(2 - f_n)^{-1}.
\end{align}

For $0 < f_n \leq 1$, this gives $1/2 < \bar{f}_n \leq 1$. 
Therefore, neglecting explicit n$\gamma$ interactions introduces at most a factor of 2 correction to the neutron component in our model. 
This cycling allows the neutral nucleon fraction to be sustained over
distances larger than a single neutron decay length. This is because
neutrons are repeatedly regenerated along the CR beam. 
Our approach is therefore conservative. It captures the existence
of a neutral channel that breaks magnetic confinement, but underestimates
the extent that neutron–proton cycling can prolong this channel and
reprocess the escaping spectrum in detail. A full treatment of this process,
including the resulting spectral reshaping of the escaping CRs, is left to
future work. 

\section{Summary and Conclusions}
\label{sec:sec5}

In this study, we have presented a first 
explicit determination of how neutron-mediated escape can regulate the exchange of 
CR baryons across the hierarchy of magnetized structures in the Universe, ranging from galaxies and clusters to cosmological filaments and their surrounding voids. 
We put focus on addressing the long-range 
transfer effects of CRs across structural boundaries brought about by neutron jumps, and how this modifies CR exchange at energies ranging from  $10^{12}\,\mathrm{eV}$ (the energy at which CRs are expected to begin to build up in large-scale environments~(see \cite{Wu2024Univ})) 
to $10^{21}\,\mathrm{eV}$ (comparable to the most energetic CR particles ever detected; see \cite{Bird1995ApJ, TA2023Sci}). Our findings are summarized as follows.   

\begin{enumerate}
    \item Ballistic jumps in CR propagation caused by the production of CR neutrons in hadronic interactions qualitatively changes the fate of injected high-energy CR particles in the large-scale structure of the Universe. In regimes where an absorption-only treatment predicts strong CR suppression, neutron production partially restores an escaping component by providing a ballistic bypass across escape surfaces. This implies that the highest-energy CRs can contribute directly to the CR population beyond their host structure, even when the charged CR primaries would otherwise be lost to p$\gamma$ processes. This neutron bypass channel is relatively more pronounced in higher-redshift environments.  
    \item In clusters and filaments, neutron-mediated escape is primarily driven by p$\gamma$ interactions. In starburst galaxies, pp interactions can be dominant, but this does not operate at sufficiently low energies to allow for the substantial escape of CRs from galaxies which would otherwise confine them.  
    \item For elongated filaments, the geometry generally favors CR escape through the filament walls rather than along the spine, resembling a ``leaky pipe'' scenario rather than an ideal CR ``highway''. This supports a picture where cosmological filaments can organize CR transport at the highest energies, but can also vent a considerable fraction of their higher-energy CR content into surrounding low-density regions, including adjacent voids.
    \item In galaxy clusters, the production of CR neutrons by p$\gamma$ interactions can sustain a partial ultra-high-energy escaping nucleon flux via a ballistic bypass of magnetic confinement (and cycling of n$\gamma$ and p$\gamma$ interactions). This relaxes any requirement that CRs must avoid strong interaction environments to escape, meaning that candidate sources of CRs embedded within galaxy clusters remain viable as a source class of the ultra-high-energy CR flux detected at Earth. In the context of the cosmic web, the same mechanism supports the efficient transfer of CR nucleons from galaxy clusters into connecting filaments, potentially bypassing magnetic barriers at \mbox{cluster--filament interfaces. }
\end{enumerate}

Despite the limitations of the simple approach adopted in this study, 
  we have obtained useful insights into how ballistic jumps can alter CR propagation at ultra-high energies when neutrons are produced. 
In particular, our results indicate a qualitative shift in 
the expected CR exchange between galaxy clusters, filaments, and voids once neutron production and their ballistic 
escape are included. 

To quantify these effects more precisely, 
   future work will require 
(i) a composition-dependent (nuclear) injection and propagation framework; (ii) 
realistic CR source rates in each environment, evolved over cosmological timescales to capture the relevant non-thermal power generation history that is supplied to large-scale structures; and (iii) a more realistic treatment of magnetic fields and interface regions, guided by future observations that can trace magnetized pathways. Together, these steps will 
allow the model predictions to be mapped more confidently onto cluster--filament--void interfaces and filament ``highway'' scenarios. 
They will also clarify which observational tests can directly probe CRs in different components of the cosmic web and allow CR transfer models to be firmly tested. With that foundation, more fundamental questions can be addressed in a meaningful way---for example, whether energetic CRs can have a measurable impact the formation and evolution of large-scale structure of the Universe.

\authorcontributions{Conceptualization, E.R.O. and K.W.; methodology, E.R.O. and K.W.; software, E.R.O.; validation, Y.I., T.F., Q.H., and H.P.H.N.; formal
analysis, E.R.O.; investigation, E.R.O., K.W. and Y.I.; resources, E.R.O.; writing---original draft preparation, E.R.O.; writing---review and editing, K.W., Y.I., T.F. and Q.H.; visualization, E.R.O.; project administration, E.R.O.; funding acquisition, E.R.O. Y.I., and K.W. All authors have read and agreed to the published version of the manuscript.} 

\funding{E.R.O. acknowledges support
from the RIKEN Special Postdoctoral Researcher Program for junior scientists. He was also supported by 
the Postdoctoral Fellowship 
  of the Japan Society for the Promotion of Science 
  (JSPS KAKENHI Grant Number JP22F22327) while at the University of Osaka, where this work was initiated.  
  K.W. is supported by a 
  UKRI-STFC SA grant awarded 
  to UCL-MSSL. 
  His visits 
  to ANU Research School of Astronomy and Astrophysics, 
  where this work was initiated, 
  were supported by 
  the ANU Distinguished Visitor award. 
Y.I. is supported by an 
  NAOJ ALMA Scientific Research Grant Number 2021-17A, 
  JSPS KAKENHI Grant Numbers JP18H05458, JP19K14772, and JP22K18277, 
  and the World Premier International Research Center Initiative 
  (WPI), MEXT, Japan. 
  T.F. is supported by JST SPRING, Grant Number JPMJSP2138. 
Q.H. is supported by a UCL Overseas Research Scholarship, a UK STFC Research Studentship, and also acknowledges the support of a Short-Term JSPS Fellowship while at The University of Osaka (Fellowship ID PE25029). 
K.W., Q. H. and H.P.H.N. 
  acknowledge support 
  from the UCL Cosmoparticle Initiative. }

\dataavailability{No new data were created or analyzed in this study. Data sharing is not applicable to this article.}

\acknowledgments{The authors are grateful to the anonymous reviewers for their critical review of this work, which substantially improved the manuscript and led to refinements in our calculations. Numerical computations for this work were carried out on the HOKUSAI Bigwaterfall2 (HBW2) supercomputer system of RIKEN (project number RB240047). Test calculations and hadronic interaction event simulations were conducted on HBW2 and the Yukawa Institute Computer Facility, Kyoto University. 
  This work made use of Astropy\endnote{http://www.astropy.org, accessed on 10 January 2026.}, a community-developed core Python package and an ecosystem of tools and resources for astronomy \citep{2018AJ....156..123A, 2013A&A...558A..33A, 2022ApJ...935..167A}, Matplotlib (v. 3.10)~\cite{Hunter2007}\endnote{https://matplotlib.org, accessed on 10 January 2026.}, and the NASA Astrophysics Data System (ADS).}  

\conflictsofinterest{The authors declare no conflicts of interest.} 







\appendixtitles{yes} 

\appendixstart
\appendix

\section[\appendixname~\thesection]{Computational Methodology and Numerical Approach}
\label{sec:appendix_a}

The stochastic integral--differential transport equations used in this work involve widely differing physical scales. They are inherently `stiff', and the stochastic component makes them particularly prone to numerical instability. As such, standard numerical schemes are not reliable to obtain robust solutions. To solve Equation~(\ref{eq:integral_eq_stoch}), we therefore developed a hybrid scheme on a numerical grid with multi-level timestepping. This evolves the deterministic and stochastic components of the system separately. In each global timestep $\Delta t$, this: (i) evolves the 
CR distribution using an operator-split deterministic update, with implicit diffusion steps in space and an implicit stiff solver for energy losses and streaming in each spatial cell; (ii)  computes the expected hadronic (pp and p$\gamma$) interaction rate in every cell and energy bin and applies a Poisson-sampled sink; and (iii) for each realized hadronic interaction, Monte Carlo-samples the energy, decay time and trajectory of the resulting secondary neutrons and, depending on whether they decay inside or outside the confining structure, either re-injects the corresponding secondary CR protons or removes them from the domain. The methodology is described in detail below.

\subsection{Deterministic Component}
\label{sec:app_a_deterministic}

To solve the deterministic part of Equation \ref{eq:integral_eq_stoch} 
within a timestep from $t_k$ to $t_{k+1}=t_k+\Delta t$, the spatial and energy coordinates are discretized on a fixed numerical grid, with energies logarithmically spaced with 50 intervals over the energy range of interest in this work, between $10^{12}$ eV and $10^{21}$ eV.\endnote{Above proton energies of $2\times 10^{19}$ eV, neutron yields from {\tt SOPHIA} are not available. For neutron production above this interaction energy, 
we anchor the spectral shape to the highest-energy results available 
and extrapolate only its overall normalization and the position of the spectral peak, with both trends measured from the highest-energy reliable {\tt SOPHIA} runs. 
A more thorough determination of the neutron spectrum at these 
ultra-high energies is needed in future studies. 
However, we have found that the exact shape of the neutron production spectrum 
in this energy regime does not have a noticeable impact on the results presented in this work (for reasonable variations).}

The spatial grid is configured according to the geometry of the solver. In the spherical case (used for galaxy clusters), a three-dimensional grid over $(r_{\rm i}, \theta_{\rm j}, \phi_{\rm \ell})$ is adopted. 
For each angular cell $(\theta_{\rm j}, \phi_{\rm \ell})$ we evolve $N(r_i,E_m,t_n)$ using an operator-split deterministic update. 
Radial diffusion is treated implicitly via a conservative finite-difference discretization 
of  \((1/r^2)\partial_r[r^2 K(r,E)\partial_r N]\) on the radial grid. After 
this spatial update, we solve the stiff energy-loss and streaming terms independently in each radial cell using the implicit solver, {\tt RADAU5}~\cite{hairer2010solving}. Internally, {\tt RADAU5} takes adaptive sub-steps with a relative tolerance
${\tt RTOL}=10^{-3}$ and absolute tolerance ${\tt ATOL}=10^{-4}$, which gives a good
balance between numerical accuracy and computational time. 

We impose a reflective boundary condition at the origin and an outflow condition at the
outer boundary of the computational domain. Numerically, this is enforced by imposing a vanishing
radial gradient at $r=0$ 
corresponding to reflective symmetry, and 
an absorbing boundary at the outer edge of the grid. Continuous (Bethe--Heitler) energy losses are discretized
on the logarithmic energy grid across three adjacent points, with modified one-sided forms adopted at
the edges of the grid to avoid introducing instability. 
CR streaming is included as an additional deterministic loss term in the energy–space solver. For each spatial cell, we compute an effective streaming velocity, include the associated sink term in the {\tt RADAU5} update and add corresponding escape events from the structure to a global escape tally. 
Advection is not explicitly included in this work. 

Primary CRs are
continuously injected within a designated source region inside the cluster, filament or galaxy radius,
with a radial weighting that follows the gas density profile. In our calculations, the
hadronic (pp and p$\gamma$) interaction losses are handled entirely by the stochastic solver component
(Appendix~\ref{sec:app_a_stochastic}), so the corresponding sink term is omitted from the deterministic 
step to avoid double-counting particle losses.

In the cylindrical case (used for the cosmological filament and galaxy applications), the numerical
set-up is analogous to the spherical case, but with diffusion operating independently in the two spatial dimensions, $(R,z)$. 
Radial diffusion is implemented in a conservative form, as $(1/R)\partial_R\left[ R\,K(R,z,E)\,\partial_R N\right]$, discretized using second-order finite differences and evaluated at cell interfaces. Diffusion along the filament axis (the $z$-direction) is treated implicitly
in a separate step after the radial diffusion update and before the energy/streaming update. We apply a finite-difference step for the anisotropic
diffusion term $\partial_z[K_\parallel(R,z,E)\,\partial_z N]$, 
with $K_\parallel$ taken to be a fixed
multiple of the perpendicular diffusion coefficient $K_{\perp}(R,z,E)$. Absorbing end-caps are implemented by 
imposing Dirichlet boundary conditions $N=0$ 
at either end of the structure in the implicit $z$-diffusion solver step. This represents the free escape of CRs through the ends of the
structure. As with the spherical calculation, the streaming escaped flux is also tallied. However, in the cylindrical set-up, this escape tally is further decomposed into side versus end escape according to the local distance to the radial and vertical boundaries. This allows us to quantify the relative importance of radial versus vertical
escape.

\subsection{Stochastic Component, Sinks and Sources}
\label{sec:app_a_stochastic}

In our calculations, 
 the terms for 
 hadronic (pp and p$\gamma$) 
 interactions   
 and their associated neutron jump 
 in Equation~(\ref{eq:integral_eq_stoch})
 are treated stochastically  
 to suppress numerical stiffness 
 and to avoid large discontinuous changes 
  in the particle density within individual grid cells. 
Explicitly including these terms 
  in the deterministic solver 
  will introduce interaction timescales 
  that are much shorter than 
  the scales of diffusion 
  and continuous losses. 
This makes the Jacobian ill-conditioned 
  and forces excessively small timesteps. 
We therefore split each global timestep 
   $\Delta t$ 
   into a deterministic update
  (see Appendix~\ref{sec:app_a_deterministic}) followed by a stochastic update 
  that accounts 
  for hadronic sinks 
  and secondary CR injection. 

For each spatial cell and energy bin, 
  we first compute the expected number 
  of hadronic interactions 
  over the global timestep 
  $\lambda_{\rm phys}(r,\gamma)\,\Delta t$ 
  from the local CR density and
the sum of the rates 
  of pp and p$\gamma$ interactions. 
To maintain numerical stability 
  when the interaction rate is large, 
  we introduce a  weighting factor 
  $V_{\rm w}$,  
  which is essentially a ``fake'' volume,  
  such that the weighted mean number 
  of interaction events, 
  $\lambda_{\rm w} \equiv \lambda_{\rm phys}\,V_{\rm w}$, 
  is kept close to a target value of 
  the order unity. 
In each cell and energy bin,
  we simulate $N_{\rm MC}$ 
 independent Monte Carlo realizations 
 of the hadronic interaction process. 
For each realization, 
  we draw an integer number 
  of interaction events,  
  $N_{\rm int}^{p}$,  
  from a Poisson distribution 
  with a mean $\lambda_{\rm w}$, 
  which is the expected number of interaction events represented by that realization over one global timestep. 
  The corresponding sink 
contribution per unit volume is 
$\Lambda_{\rm sink}^{p} = -N_{\rm int}^{p}/V_{\rm w}$, and the net 
sink density applied to the post-deterministic solution is obtained by 
averaging over events, 
$\Lambda_{\rm sink} = N_{\rm MC}^{-1} \sum_p \Lambda_{\rm sink}^{p}$. 
By construction, $\langle \Lambda_{\rm sink}\rangle = -\lambda_{\rm phys}$, 
so that the expectation value matches the physical hadronic interaction rate, while the 
individual realizations remain numerically manageable, even when a large 
fraction of the particles in a grid cell are removed in a single 
timestep. 

Secondary neutron production 
  and their energy distribution  
   is determined using precomputed production kernels. 
For each combination 
  of primary CR proton energy 
  and secondary CR neutron energy, 
  we tabulate the energy-dependent 
multiplicity and event rates for pp interactions in ambient gas, and p$\gamma$ interactions 
  in the CMB, dust-reprocessed radiation 
  and stellar radiation fields 
  (using {\tt SOPHIA} and {\tt AAfrag}). 
In the stochastic 
update, each Monte Carlo interaction realization is 
interpreted as producing a bundle of secondary neutrons.
For each bundle,  
  we sample a neutron Lorentz factor $\gamma_{\rm n}$ from 
the tabulated neutron energy distribution at the parent proton energy, 
and then draw a decay time from the exponential distribution $t_{\rm dec} = -\gamma_{\rm n}\,\tau_n \ln u$ 
for a uniform random variate $u\in[0,1]$. 
The propagation direction of the neutrons is expected to almost exactly follow the direction 
of their parent protons. Although our calculation approach does not track individual particles, we emulate this directional inheritance by assigning 
a neutron propagation direction that is collinear 
with the local primary proton flux gradient. The neutron is then 
propagated in a straight line at speed $c$ to its decay point. 

We next determine whether the decay occurs within or outside the confining structure. If the decay lies outside the surface of the structure, the 
corresponding converted CR bundle is counted as having escaped. In the cylindrical configuration, when a neutron leaves the confining magnetized structure, we also record whether it exited through the cylindrical side or through one of the end-caps, based on the first intersection of its trajectory with the escape surface.  
If the neutron decays inside the magnetized domain, we map the decay 
location back to the nearest grid cell and treat the decay products as 
freshly injected secondary CR protons. The secondary source term in that 
cell is then computed by rescaling to the realized number 
of interactions in that cell and timestep. This yields a stochastic 
secondary injection rate $q_{\rm sec}(E_{\rm s})$, which is added to the 
CR distribution in the decay cell.  
This ensures that fluctuations in 
the secondary population remain self-consistent with the Poisson-sampled 
number of primary interactions in the originating cell. 

This scheme is designed to obtain quasi-steady solutions over many global 
timesteps. 
It is appropriate for the applications
considered here since the evolutionary timescales of galaxy clusters,
cosmological filaments and galaxies are many orders of magnitude longer
than the CR interaction and transport timescales at the energies of 
interest. For very short-duration or strongly time-dependent scenarios,
a much smaller global timestep or a fully coupled time-dependent
treatment of transport and hadronic interactions would be required to
obtain accurate solutions. 

\begin{adjustwidth}{-\extralength}{0cm}
\printendnotes[custom] 

\reftitle{References}

\PublishersNote{}
\end{adjustwidth}
\end{document}